\documentstyle[12pt,epsf]{article}
\newcommand{\be}{\begin{equation}}
\newcommand{\ee}{\end{equation}}

\begin{document} YITP-00-51 
\hspace{10cm}
\today
\\
\vspace{3cm}
\begin{center} {\LARGE   Mode regularization of the susy sphaleron and kink:\\

 zero modes and discrete gauge symmetry   } 
\\  \vspace{2cm} {Alfred Scharff Goldhaber\footnote{e-mail:
goldhab@insti.physics.sunysb.edu},  Andrei Litvintsev\footnote{e-mail:
litvint@insti.physics.sunysb.edu}  and Peter van
Nieuwenhuizen\footnote{e-mail: 
vannieu@insti.physics.sunysb.edu}\\
 { \it C.N.Yang Institute for Theoretical Physics, SUNY at Stony Brook},
\\ {\it Stony Brook, NY 11794 } }
\abstract{    To obtain the one-loop corrections to the mass of a kink
by  mode
regularization, one may take one-half the result for the mass of a widely
separated kink-antikink (or sphaleron) system, where the two bosonic
zero modes
count as two degrees of freedom, but the two fermionic zero modes as
only one
degree of freedom in the sums over modes. For a single kink, there is
one bosonic
zero mode degree of freedom, but it is necessary to  average over four
sets of
fermionic boundary conditions in order (i) to preserve the fermionic
Z$_2$ gauge invariance $\psi \to -\psi$, 
(ii) to satisfy the basic principle of mode regularization that the boundary conditions in the trivial and the kink sector
should be the same,
(iii) 
in order that the energy stored at the boundaries cancels and
 (iv)  to avoid obtaining
a finite,
uniformly distributed energy which would violate cluster decomposition.  The
average number of fermionic zero-energy degrees of freedom in the
presence of the
kink is then indeed 1/2. For boundary conditions leading to only one fermionic
zero-energy solution, the Z$_2$ gauge invariance identifies two
seemingly distinct
`vacua' as the same physical ground state, and the single fermionic zero-energy
solution does not correspond to a degree of freedom. Other boundary conditions
lead to two spatially separated $\omega \sim 0 $ solutions,
corresponding to one
(spatially delocalized) degree of freedom.  This nonlocality is
consistent with
the principle of cluster decomposition for correlators of observables.
 }
\end{center}
\newpage

\section{Introduction}

The problem of how to compute the one-loop corrections  to the mass $M$ and central
charge $Z$ of supersymmetric (susy) kinks has been the subject of renewed
investigations in the  past few years.  In this article we give a precise
prescription for computing the mass of ordinary and susy kinks using mode
regularization. The prescription follows from a careful study of bosonic and
fermionic zero modes.  Differing from previous prescriptions, it yields the
accepted result, thus reaffirming mode regularization as a bona fide scheme.

According to standard arguments \cite{raj}, the mass can be written in
terms of
differences of sums over zero-point energies, but because the sums
themselves are
divergent, one must  specify how to regularize them. Furthermore, the
number and
values of the zero-point energies depend on the boundary conditions one imposes
on the fields, and hence one must decide which boundary conditions to
use, or,
more precisely, when and which parts to subtract from the sums over zero-point
energies for a given set of boundary conditions.  When boundary conditions
distort the fluctuations at the boundaries, one must first subtract the extra
boundary energy to obtain the mass corrections for the kink  and superkink.
In addition, as we shall show, certain boundary conditions lead to a uniformly distributed energy density, 
which also should not be counted as part of the kink mass.

The recent interest in the subject of this paper began with the work \cite{rebhan}, where two important questions
were posed:

\noindent
1)  What is the quantum correction $M^{(1)}$ to the mass of the kink in
supersymmetric $\lambda \phi^4$ theory?

\noindent
2)  If  $M^{(1)}$ is not zero, can the kink remain a BPS-saturated state
in the
presence of quantum corrections, and if so how does this occur?

The authors of \cite{rebhan} noted that the two regularization
methods they used to calculate  $M^{(1)}$, momentum cutoff and mode
cutoff, gave
different answers. Moreover, the BPS bound did not seem to be saturated. Since then
various regularization schemes have been applied to this problem:
momentum cut-off \cite{rebhan,misha,we}, mode cut-off \cite{rebhan,we},
mass-derivative regularization \cite{misha,we},  phase-shift methods \cite{graham},
higher-derivative supersymmetry-preserving regularization \cite{shifman}, dimensional regularization
\cite{shifman} and derivative expansion \cite{dunne}. 

From \cite{we} one sees that the inadequacy in the momentum
cutoff calculation had to do with the need for smoothing the cutoff so
that it
becomes well-defined.  In the present work we shall show that the mode cutoff
calculation was correct, but included a localized boundary energy along
with the
kink energy.

Nastase, et al. \cite{misha} avoided these pitfalls by first evaluating the mass-derivative of the mode sums, 
which gave better control of the divergences
and thus
eliminated the need for smoothing the cutoff.  Their result for  $M^{(1)}$
agreed with the older work of Schonfeld \cite{schonfeld} for the kink-antikink
system, suggesting that this indeed is the correct value.  They also suggested
that there might be an anomaly which would restore the BPS condition.  

The MIT group \cite{graham} used
continuum phase shift methods (avoiding consideration of boundary
conditions) to
compute the one-loop corrections to the kink energy  $M^{(1)}$ and to the central
charge 
$Z^{(1)}$, finding that they are the same, so that the BPS condition is
obeyed.  They did not ascribe the shift in $Z$ to an anomaly, instead
treating it as a straightforward one-loop result. 

The Minnesota group \cite{shifman}, stimulated by \cite{rebhan} and \cite{misha},
undertook to attack the second question of \cite{rebhan} directly. 
There is a
beautiful argument originated by Witten and Olive \cite{WO}.  
If the isolated kink has a single ground state, and
if perturbative quantum corrections violated the condition $M=Z$,
then the fact that this would mean the supersymmetry is completely broken
implies that there must be a double degeneracy of the ground state,
hence a
multiplicity discontinuity at $\hbar = 0$.  In \cite{shifman} three
different methods were used to calculate the one-loop correction to $Z$, all
methods agreeing that the correction represents a local anomaly, and results
in maintaining  $M=Z$, with $M$ given by (\ref{ura1}). The authors of \cite{shifman} 
also computed the one-loop effective superpotential, 
and extracted from it the effective central charge
as the difference of the effective superpotential at $\pm \infty$; this yielded the one-loop correction to $Z$. 

In \cite{we}, the anomaly in the central charge was directly computed using momentum cutoff
to
regulate both the Dirac delta function appearing in the algebra of the supersymmetry
charges and the propagators appearing in the loops; again  $M=Z$ was found.  

Perhaps the
easiest scheme
is mass-derivative regularization 
\cite{misha}, according to which one first evaluates the derivative 
$\frac{\partial}{\partial m}M=\frac{\partial}{\partial m} \left[ \frac{1}{2} \sum \omega + \Delta M \right]$
of the
sums (which is better convergent, so that there is no sensitiviy to the
form of
cutoff) and then integrates w.r.t.
$m$ using the renormalization condition $M(m=0)=0$.  Any boundary
conditions on
the fluctuations for which the divergences in $M^{(1)}$ cancel are
allowed.  
(There are boundary conditions for which $M^{(1)}$ diverges, for example susy
boundary conditions. The  divergence is then a boundary effect which
must be
subtracted by hand.) A detailed discussion is given in
\cite{we}; here we only need the result: the one-loop bosonic, fermionic and
supersymmetric corrections to the kink mass are, respectively,
\be 
\begin{array}{c} M^{(1)}_b =  -m \hbar \left( \frac{3}{2 \pi} -
\frac{\sqrt{3}}{12} \right); \hspace{1cm} M^{(1)}_f =  m \hbar \left( \frac{1}{
\pi} -
\frac{\sqrt{3}}{12} \right); 
\\ \\
 M^{(1)}_s = M^{(1)}_b +  M^{(1)}_f =  - \frac{m \hbar}{2 \pi} \\ 
\end{array}
\label{ura1}
\ee These values are now accepted by all workers in the field [3-7]. 

The problem to be solved is thus how to obtain these results with the other
regularization schemes and  other boundary conditions. The most commonly used
schemes are energy cut-off (= momentum cut-off) and mode regularization. Although
each has been in use for decades, we claim that each needs
modifications. For
energy cut-off regularization ( in which one first computes each of the
sums up
to the same given energy $\Lambda$, and then takes the limit $\Lambda \to
\infty$) it was found that a simple modification makes the sum well
defined, and
reproduces
 (\ref{ura1}): instead of an abrupt cut-off at $\Lambda$ one needs a smooth
cut-off which interpolates between the zero-point energies in the
topological and
trivial sectors \cite{we}. 

In this article we repair mode number regularization. The basic idea of this
scheme  is to subtract an equal number $N$ of modes
$\omega_n^{(0)}$ in the trivial sector from the modes $\omega_n$  in the
topological sector, and then to take the limit $N \to \infty$,  but a problem
arises whether one should include some or all or none of the zero  modes
in this
counting. This issue has turned out to be surprisingly complicated, and for
pedagogical reasons we shall first deduce in section 2 the correct rules by
requiring that they reproduce the result in (\ref{ura1}). The problem then
obviously is to justify these rules. We shall first consider a kink-antikink
configuration, which lies in the trivial sector (having no overall winding
number) so that standard manipulations of quantum field theory are still
reliable.  For such a system  the energy located at the boundaries in the
kink-antikink sector cancels  the same quantity in the trivial sector if
one uses
the same boundary conditions  in both sectors\footnote{
If one first considers a finite number of modes in the sector where the classical
scalar field is constant, and then slowly turns on the kink-antikink 
configuration by pulling the scalar field around $x=0$ away from its constant value, 
the mode energies move from their values in the trivial
background  to their values in the $\bar{K}K$ configuration.  This is the justification
for mode regularization, as in (\ref{sum0}). Since a change in the background
away from the boundaries will not change the (localized or delocalized) boundary
energy, the latter (if present) cancels between the $\bar{K}K$  case
and the
trivial case.}, so the mass of the kink
is then
just one-half of the sums over the zero-point energies plus the counter
term for
mass renormalization (the latter will be given in section 2). (For
periodic or
antiperiodic boundary conditions in the kink-antikink system, there is
not even
any localized boundary energy because these boundary conditions are translationally invariant). 
For the bosonic case one finds for the
kink mass (putting $\hbar=1$)
\be 
\label{sum0} M^{(1)}_b=\frac{1}{2} \lim_{N \to \infty} \sum_{n=1}^N (\frac{1}{2}
\omega_n^{b}-\frac{1}{2} \omega^{b,(0)}_n) + \Delta M_{b} 
\ee   where $\frac{1}{2} \omega_n^b$ are the zero point energies for the bosonic
fluctuations around the kink--antikink background, $\frac{1}{2}\omega_n^{b,(0)}$
those around the trivial background, and  $\Delta M_b$ is the
counterterm for a
single kink.

Next we consider the susy kink. Except for a unique value of the
strength of the
Yukawa coupling  of the fermions, the susy of the action is explicitly and
completely broken, but in the $\bar{K}K$ (antikink--kink) background
zero modes
remain. We shall therefore generalize our approach and consider
arbitrary kinks
with fermions, and not only susy kinks. The action we use contains a
Yukawa term
$ - c \sqrt{\frac{\lambda}{2}} \phi \bar{\psi} \psi
$  with $\psi$ a real two-component spinor and $c=1$ for susy. The
correction to
the mass of the kink is given by
\be
\label{sum1} M^{(1)}_s=\frac{1}{2} \lim_{N \to \infty} \sum_{n=1}^N [ \frac{1}{2}
\omega_n^{b} - \frac{1}{2} \omega_n^{b, (0)} - \frac{1}{2}  \omega_n^{f} +
\frac{1}{2} \omega_n^{f, (0)} ] +
\Delta M_{[b+f]}
\ee where $b$ denotes bosonic frequencies and $f$ fermionic ones and the
counterterm $\Delta M_{[b+f]} = \Delta M_{b}+ \Delta M_{f} $  is due to both
bosonic and fermionic loops.

There are various ways to describe a kink-antikink background. By far
the most
used is  the configuration which describes a kink centered at $L/2$ for
$x \ge 0$ and an antikink centered at $-L/2$ for $x<0$.   
\\
\epsfbox{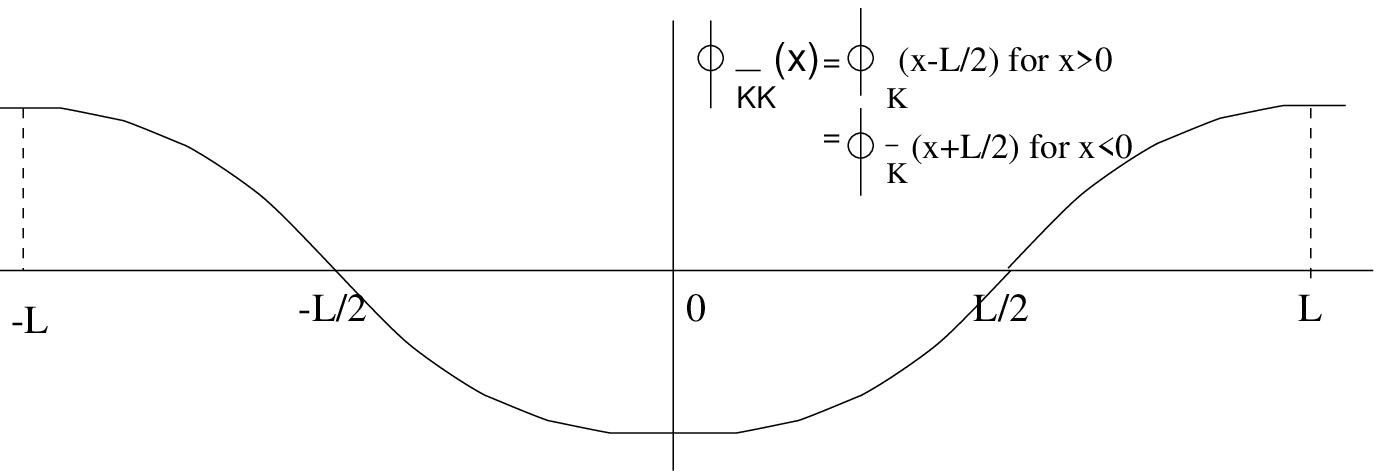}  \vspace{0.3cm}
\\ Figure 1. {\it The kink-antikink configuration.} \\ \\ This
configuration has
the slight drawback that at $x=0$ the field $\phi(x)$ is not
differentiable (the
left- and right- derivatives differ),  so that $\phi(x)$ is not a
solution of the
field equation. We shall call this configuration
$\phi_{\bar{K}K}(x)$. Another configuration one might consider is everywhere
differentiable, but nowhere a solution of the field equations:
$\phi_{K+\bar{K}}(x)=
\phi_K(x) + \phi_{\bar{K}}(x)+ \frac{\mu}{\sqrt{\lambda}}$. For large  positive
$x$ the sum of $\phi_{\bar{K}}(x)$ and $ \frac{\mu}{\sqrt{\lambda}}$
vanishes and
one obtains the usual kink solution $\phi_K(x)$, while for large
negative $x$ one
is left with $\phi_{\bar{K}}(x)$.  However, it is difficult to determine the
spectrum of fluctuations around this background, and we shall not use it
below. A
third configuration one might consider is one of the sphaleron solutions
$\phi_{sph}(x)$ of \cite{manton}, which are defined on the interval $-L
\le x \le L$ with the periodic boundary conditions
$\phi_{sph}(-L)=\phi_{sph}(L)$ and $\phi^\prime_{sph}(-L)=\phi^\prime_{sph}(L)$. (The sphaleron is thus defined on a
circle). For
our purposes the sphaleron solution which becomes one kink-antikink pair
as $L$
tends to infinity is the relevant one, and we shall hereafter refer to
it as `the
sphaleron'. We shall see that our results for mode regularization  are
the same
for the $\bar{K} K$ background as for the sphaleron background.

Our main conclusion for the bosonic kink-antikink system, to be derived
below, is
that both translational zero modes should be taken into account in the
sum over
zero-point energies (\ref{sum0}). When the kink and antikink are not infinitely
far apart and are described by the sphaleron background, one still has a zero
mode with $\omega=0$ for translations, while a second mode has
$\omega^2<0$ and indicates an instability  (the kink is attracted by the
antikink). For
$\omega=0$, one has the usual collective coordinate $\hat{X}$ (the Hamiltonian does not depend on
$\hat{X}$) for
translations and
its canonically conjugate momentum $\hat{P}$, so that $\hat{X}$ and $\hat{P}$
form a canonical pair and correspond to one term in the sum over
$\omega_n^b$ in (\ref{sum0}). For
$\omega^2<0$ the solutions with $e^{\pm |\omega| t}$ define one pair
 of canonical variables and thus another  term in the sum over zero-point
energies. (The Hamiltonian depends in this case on both canonical
variables). In
the
$\phi_{\bar{K}K}$ background there is one pair of solutions with $\omega^2<0$,
but the other pair of solutions now has $\omega^2>0$. As usual the
solutions with
$e^{\pm i \omega t}$ then describe one degree of freedom. Thus there are
still two
degrees of freedom associated with the bosonic zero modes. 

For fermions the situation is quite different. 
For the sphaleron
background we
find two fermionic solutions with $\omega^2=0$ (see (\ref{tufta18})). The corresponding operators
$\beta$ and $\gamma$ satisfy $\beta^2=\gamma^2=1$ and $\{ \beta, \gamma
\}=0$  and thus again there is one degree of freedom.
Next we
study the
$\phi_{\bar{K}K}$ system with a finite separation $L$. Then we find two
solutions of the Dirac equation with the same (very small) $\omega^2>0$
which are
normalizable and which enter in the second quantization of the fermionic
fluctuation field
$\psi(x,t)$ (for $x>0$)  as follows
\be
\psi(x,t) = b \left( \begin{array}{c} \psi_K(x) \\ \frac{i}{\omega} (\partial_x+c
\sqrt{2 \lambda} \phi_K ) \psi_K(x)  \end{array}
\right) e^{-i \omega t} + b^\dagger \left( \begin{array}{c} \psi_K(x) \\
-\frac{i}{\omega} (\partial_x+c \sqrt{2 \lambda} \phi_K ) \psi_K(x)  \end{array}
\right) e^{i \omega t} 
\ee Here $\psi_K(x)$ is a real normalizable function, $\phi_K(x)$ is the kink solution and $c=1$ for the
susy case.
Clearly $b$ and $b^\dagger$ form {\bf one} conjugate pair, hence one
degree of
freedom.  Therefore the final effect of fermionic zero modes amounts to
one term
in the sum over $\omega_n^f$ in (\ref{sum1}). 

Moving the kink and antikink apart, one obtains  a free kink and a free antikink,
each having its own zero mode in the Dirac equation (even for $c \ne
1$). The 
problem then is how to perform mode number regularization for a single isolated
kink. As we shall discuss later, for a canonical description  one must
take into
account four sets of boundary conditions, and average the results.  
Then the
fermionic ``half degree of freedom'' of the kink  appears  as a change
in degrees
of freedom (from vacuum) by unity in one pair of boundary conditions,
and no
change at all in the other pair. 

Closer inspection reveals that for certain boundary conditions there is exactly
one fermionic zero-energy solution.  As the ground state then is an eigenstate of the
operator which appears as the coefficient of this solution in the
fermion field, 
 the Fock space for this system is half as big as one might have expected,
meaning that by this elementary criterion there is no zero-energy
fermion degree
of freedom.

In string theory one encounters a similar situation with respect to the
zero mode
of the coordinate ghost (denoted by $c_0$). In that case BRST cohomology shows
that states with $c_0$ are BRST exact, so that there is no doubling of
the number
of states. In our case we do not have BRST symmetry to remove half of
the states
of the Fock space, but we shall show that one can divide Fock space into two
sectors, such that all operators map states of one sector into states of
the same
sector.  The other sector is then a Z$_2$-gauge copy of the first, and
as a
discrete Z$_2$-gauge symmetry in string theory can be promoted to a continuous
symmetry, a BRST approach may be possible also for the susy kink. 

\section{The correct rules for mode regularization}

In this section we demonstrate that for the kink-antikink system 
counting two
zero modes in the bosonic spectrum but only one in the fermionic 
spectrum gives
the correct answers (\ref{ura1}) for both the susy and the bosonic case.
In the
next sections we derive these rules. After a brief review of the
properties of
the spectra, the actual calculation of the mass is performed in
(\ref{s99})  and
(\ref{k99}).

The Lagrangian is given by 
\be {\cal L}= -\frac{1}{2} (\partial_\mu \phi)^2 - \frac{1}{2} U^2(\phi) -
\frac{1}{2} \bar{\psi} \gamma^\mu \partial_\mu \psi - \frac{1}{2} c
\frac{dU}{d \phi} \bar{\psi} \psi
\label{Lagrangian}
\ee  where for the kink $U[\phi(x)] = \sqrt{\frac{\lambda}{2}} \left(
\phi^2 - 
\frac{\mu_0^2}{\lambda} \right) $ and $\bar{\psi}=\psi^{\dagger}  i
\gamma^0$. We shall use the representation $\gamma^1=\left(
\begin{array}{cc}  1 & 0
\\ 0 & -1
\\
\end{array} \right)$ and $\gamma^0=\left( \begin{array}{cc}  0 & -1 \\ 1
& 0 \\
\end{array} \right)$ and take $\psi$ real.   This system has susy when
$c=1$, and then the susy transformation rules are $\delta \phi= \bar{\epsilon}
\psi$ and
$\delta \psi = \gamma^\mu \partial_\mu
\phi \epsilon - U \epsilon$. The theory with or without fermions is renormalized
by replacing $\mu_0^2$ by $\mu^2+\delta \mu^2$, where for the susy case $
\left( \delta \mu^2 \right)_{s} = 
\frac{\lambda \hbar}{4 \pi} \int_{-\Lambda}^{\Lambda}
\frac{dk}{\sqrt{k^2+m^2}} $ (we put $m^2=2 \mu^2$).  For the bosonic
case  $
\left( \delta \mu^2 \right)_{b} = 
\frac{ 3 \lambda \hbar}{4 \pi} \int_{-\Lambda}^{\Lambda}
\frac{dk}{\sqrt{k^2+m^2}} $. The one-loop correction to the mass of the
supersymmetric kink is given by (\ref{sum1}) with counter-term $\Delta
M_{[b+f]}=\frac{m}{\lambda} (\delta
\mu^2)_{s}$. For the bosonic case we use (\ref{sum0}) with  $\Delta
M_{b}=\frac{m}{\lambda} ( \delta \mu^2)_{b}$. The value of $\delta
\mu^2$ follows from requiring absence of tadpoles and the value of
$\Delta M$
follows from replacing $\phi $ by $\phi_K$ in $\int_{-\infty}^{\infty}
\frac{1}{2} U^2 (\phi) dx$ and retaining the term linear in $\delta
\mu^2$. For details see \cite{rebhan}.

Expanding the bosonic field around the background configuration $\phi(x,t)=
\phi_{\bar{K}K}(x,t)+\eta(x) e^{-i \omega t}$ one finds the following equation
for the bosonic modes $\eta$:
\be
\label{bof}
\label{uf6}
\omega^2  \eta + \partial_x^2 \eta - [(U^\prime)^2 + U^{\prime \prime}U] 
\eta = 0
\ee The solutions of this equation for the single kink background can be found
explicitly (see, for example, \cite{raj}). The spectrum consists of  a translational zero mode, a
bound state
$\omega_B=\sqrt{3}m/2$, and a continuum of
states. 
From this information one can easily extract the spectrum of the kink-antikink
configuration at large $L$.

For the fermionic fields we set $\psi(x,t)=\left( \begin{array}{c}
\psi_1(x) \\
\psi_2(x) \end{array} \right) e^{-i \omega t}$ and obtain the Dirac
equation  
\be
\label{dirac1}
\begin{array}{c} i \omega \psi_2 + ( \partial_x + U^\prime) \psi_1 = 0
\\ - i
\omega \psi_1 + ( - \partial_x + U^\prime) \psi_2 = 0 \ \ . \\
\end{array}
\ee

For $\omega=0$, the single kink (or antikink) in a box defined by $-L/2
\le x \le L/2$  has two fermionic solutions with periodic boundary conditions,
namely the  zero mode attached to the kink, obeying
\be
\begin{array}{c}
\label{dirac4}
\partial_x \psi_1 + U^\prime \psi_1 = 0 \ \ , \\
\end{array}
\ee    where $U^\prime= m \tanh \frac{mx}{2}$, and a second zero mode
attached to
the boundary, obeying
\be
\begin{array}{c}
\label{dirac4'}
\label{uf9}
\partial_x \psi_2 - U^\prime \psi_2 = 0 \ \ .\\
\end{array}
\ee
 The solutions to these equations are
\be
\label{uf10}
\label{mody0}
\psi_I = \left(
\begin{array}{c}
 {a_1}/{\cosh^{2} \frac{mx}{2}} \\  
 0
\end{array}
\right) \ \ , 
\hspace{3cm}
\psi_{II} = 
\left( 
\begin{array}{c} 0 \\ a_2 \cosh^2 \frac{mx}{2} 
\end{array}
\right) .
\ee (for $c \ne 1$ the power $2$ becomes $2c$).

For $\omega \ne 0$ one may express each component of $\psi$ in terms of
the other,
in which case (\ref{dirac1}) becomes
\be
\label{dirac2}
\begin{array}{c}
\omega^2 \psi_1 + \partial_x^2 \psi_1 - (U^\prime)^2 \psi_1 + U^{\prime \prime}
\partial_x \phi \psi_1 = 0 \\
\omega^2 \psi_2 + \partial_x^2 \psi_2 - (U^\prime)^2 \psi_2 - U^{\prime \prime}
\partial_x \phi \psi_2 = 0 \ \ .\\
\end{array}
\ee For the case of a kink one may use the Bogomol'nyi equation $\partial_x
\phi = -U$, and (\ref{dirac2}) becomes 
\be
\label{dirac12}
\begin{array}{c}
\omega^2  \psi_1 + \partial_x^2 \psi_1 - [(U^\prime)^2 + U^{\prime \prime}U]
\psi_1 = 0 \\
 \omega^2  \psi_2 + \partial_x^2 \psi_2 - [(U^\prime)^2 - U^{\prime \prime}U]
\psi_2 = 0 \ \ .\
\end{array}
\ee These are Schr\"odinger-type equations, and the first of them is the
same as
(\ref{bof}). For the antikink sector the Bogomol'nyi equation reads 
$\partial_x \phi = U$, so one must exchange 
$\psi_1$ and $\psi_2$ in (\ref{dirac12}) keeping $U$ unchanged.

In general, the equations (\ref{bof}) and (\ref{dirac12}) for a
kink-antikink 
system on $-L \le x \le L$ can be written as
\be
\omega^2  f(x) + \partial_x^2 f(x) - V(x) f(x) = 0
\ee The potentials $V(x)$ for bosonic and fermionic fluctuations are
sketched in
Fig. 2.
\\ \\ 
\epsfbox{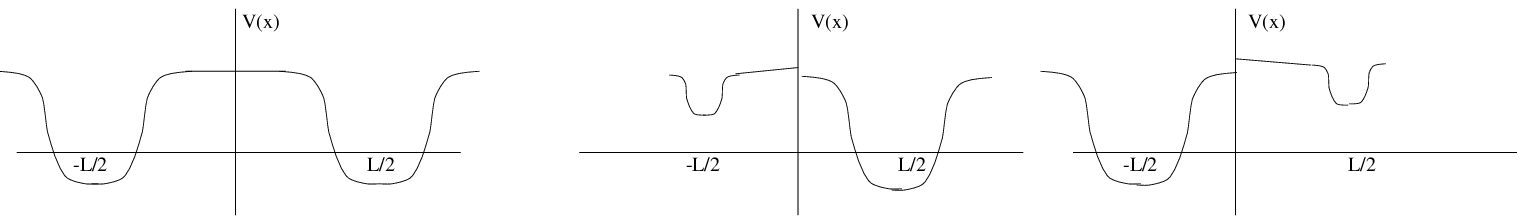} 
  
\hspace{1cm} $V(x)$  for $\eta$ \hspace{3cm} $V(x)$ for $\psi_1$ \hspace{3cm}
$V(x)$ for $\psi_2$ \\ \\ Figure 2. {\it Potentials for the bosonic and fermionic
fluctuations in  the kink-antikink system.} \\ \\ A plane wave incident
from the
right acquires a phase shift $\delta(k)$ in the deeper potential
$V(x)=\frac{1}{2} m^2 (3 \tanh^2 \frac{mx}{2}-1)$ and a phase shift
$\delta(k)+\theta(k)$ in the shallower potential $V(x)=\frac{1}{2} m^2 (\tanh^2
\frac{mx}{2}+1)$. Thus on the far right
$\psi_1 \sim e^{i[kx+\delta(k)+\frac{1}{2} \theta (k)]}$, on the far
left 
$\psi_1 \sim e^{i[kx-\delta(k)-\frac{1}{2} \theta (k)]}$ while near
$x=0$ one has
$\psi_1 \sim e^{i[kx+\frac{1}{2} \theta (k)]}$. The phase shifts are
given in
Fig. 3
\\        
$$ \epsfbox{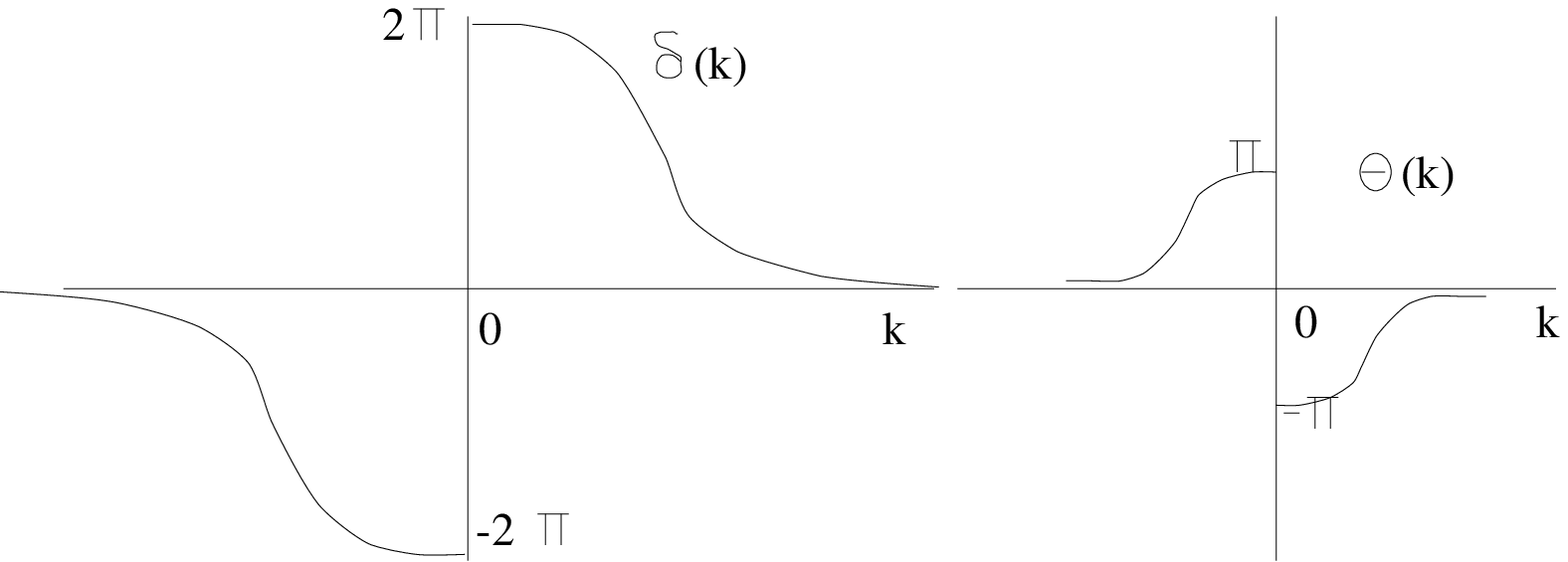} $$
$$
\begin{array}{c}
\delta (k) = 2 \pi \Theta(m^2-2k^2)- 2 \arctan \left( 
\frac{3 m k }{m^2-2 k^2}  \right) 
\hspace{0.3cm} {\rm for} \hspace{0.3cm} k>0 \\
\theta(k) = - 2 \arctan \frac{m}{k}  \hspace{0.3cm} {\rm for } 
\hspace{0.3cm} k>0
\end{array} $$ 
\\  Figure 3. {\it The phase shifts for the bosonic and fermionic fluctuations.}
\\ \\ The phase shifts are defined up to $2 \pi$. Different  authors use
different expressions for these phase shifts, which nevertheless lead to
the same
answers. 

For definiteness we choose a particular set of boundary conditions
although for the kink-antikink system to which we now turn, the results do not depend on which choice
one makes. We choose the boundary conditions of \cite{schonfeld}, 
 $\psi_1(L)=\psi_1(-L)=\eta(L)=\eta(-L)=0$ where $\eta(x)$ are the fluctuations
of the bosonic field around the background (the background is constant
in the
trivial sector,  and equal to the kink-antikink combination of Fig. 1 in the
nontrivial sector).  These boundary conditions are susy if one considers
a kink
background but they are not susy for the kink-antikink background,
because the
kink background breaks half of the susy while the antikink background
breaks the
other half. The  boundary conditions lead to the following quantization rules
\be
\begin{array}{c} k^b_{ n } L + \delta(k^b_{n}) = \frac{\pi n}{2} \ \ \ 
{\rm for
\ bosons} \\ k^f_{m } L + \delta(k^f_{ m})+ \frac{1}{2} \theta(k^f_{ m}) =
\frac{\pi m}{2}  \ \ \  {\rm for \ fermions}   \\
\label{ur3}
\end{array}
\ee where the integers $n$ and $m$ are non-negative. Solutions of
equations 
(\ref{ur3}) only exist for $n \ge 4$ and $m \ge 3$. Clearly $k_n^b=k_n^{b,0}-
\delta(k_n^{b,0})/L+ {\cal O }(1/L^2)$, with a similar expression for
$k_n^f$. 

For large $k$, the bosonic and fermionic levels with the same $m$ and $n$
approach each other, but  the bosonic energy is always a bit smaller
than the
fermionic energy. For any given  $n=m$ there is still a small interval
of momenta
such that if we would pick a cut-off in this interval, then the corresponding
bosonic level will be below the cut-off, and the corresponding fermionic level
above the cut-off. In ref. \cite{schonfeld} Schonfeld  computes
(\ref{sum1}) and
he implicitly excludes  this possibility as non-generic, having very small
probability in comparison with the chance to include both corresponding levels
(see below his eq. (2.44) ).  With this prescription, Schonfeld's procedure
becomes in effect a mode cut-off method, with one more fermionic than bosonic
mode in the continuous spectrum.

We are now ready to apply our counting rules. Let us start with the
bosonic case.
We include two zero modes in our counting. The bosonic mass correction
for  the
kink-antikink system is given by (reinserting $\hbar$)
$$ 2 M^{(1)}_{b} = \left( 2 \times 0 + 2 \frac{\hbar}{2} \omega_B  +
\frac{\hbar}{2}
\sum_{n=4}^N \sqrt{(k_n^{b})^2+m^2} \right) - \frac{1}{2}
\sum_{n=0}^N  \hbar
\sqrt{(k_n^0)^2+m^2}+ 2 \Delta M_{b}
$$
$$ = \hbar \omega_B - 2 m \hbar  + \frac{1}{2} \hbar \sum_{n=4}^N \left(
\sqrt{(k_n^{b})^2+m^2} - \sqrt{(k_n^0)^2+m^2} \right) + 2 \Delta M_{b}
$$
$$ =\hbar \omega_B - 2 m \hbar  + \frac{1}{2} \hbar \int_0^\Lambda 
\frac{dk}{(\pi /2)} \frac{d \omega}{dk} (-\delta(k)) + 2 \Delta M_{b}
$$ 
\be = \hbar \omega_B - 2 m \hbar  - \hbar \frac{\omega \delta(k)}{\pi}
|_0^\Lambda + \left[ \hbar
\int_0^\Lambda \frac{dk}{\pi} \omega \delta^\prime (k) + 2 \Delta M_{b} 
\right] = \hbar \omega_B - \frac{3 m \hbar}{\pi}  + 2 \left[ - \frac{\sqrt{3}m
\hbar}{6} \right]
\label{s99}
\ee We used $\omega \delta (k) = 2 \pi m$ at $k=0$ and $\omega
\delta (k) = 3 m$ for $k \to \infty$,  while $\hbar \int_0^\Lambda \frac{dk}{2
\pi} \omega
\delta^{\prime} (k) + \Delta M_{b}= - \frac{\hbar \sqrt{3} m}{6}$ (see
\cite{rebhan}, eq (14) ). With
$\omega_B = \frac{\sqrt{3}}{2}m$ we indeed get (\ref{ura1}).  Note that
if we
would forget about  the unstable mode, i.e. if we would take only one
zero mode
into account, the result for $ M^{(1)}_{bos} $  would be divergent.  Also
omitting to include any zero mode in (\ref{sum0}) yields a divergent answer.

Next consider the supersymmetric case. The bosonic and fermionic
contributions in
the trivial sector cancel. We know already that we need two bosonic zero
modes. 
We claim that one should take only one fermionic zero mode into account (for
reasons to be explained later). The mass correction of the susy kink-antikink
system is then given by
$$2  M_{s}^{(1)} = \left[ 2 \times 0 + 2 \frac{1}{2} \hbar \omega_B +  
\frac{1}{2} \sum_{n=4}^N  \hbar \sqrt{(k^b_n)^2+m^2} \right]
$$
$$ - \left[ 0 +  2 \frac{1}{2} \hbar \omega_B   + \frac{1}{2}
\sum_{n=3}^N \hbar
\sqrt{(k^f_n)^2+m^2 }  \right] + 2 \Delta M_{[b+f]}
$$
$$  = -\frac{1}{2} m \hbar + \frac{1}{2} \hbar \sum_{n=4}^N \left(
\sqrt{(k^b_n)^2+m^2}-\sqrt{(k^f_n)^2+m^2 } \right) + 2 \Delta M_{[b+f]} 
$$
$$ = -\frac{1}{2} m \hbar  + \frac{1}{2} \hbar \int_0^\Lambda
\frac{dk}{(\pi / 2)} 
\left(  
\frac{d}{dk} \sqrt{k^2+m^2} 
\right)
\left(
\frac{1}{2} \theta (k)
\right) + 2 \Delta M_{[b+f]}
$$
\be = -\frac{1}{2} m \hbar  + \frac{\hbar}{2 \pi} \omega \theta(k)
|_0^\infty  =
- \frac{1}{2} m \hbar  - \frac{m \hbar }{\pi} + \frac{1}{2} m \hbar  = 
 - \frac{m \hbar }{\pi}
\label{k99}
\ee We used $k_n^b-k_n^f = \frac{1}{2L} \theta(k_n) + {\cal O}(1/L^2)$, and
$ -\int_0^\Lambda \frac{dk}{4 \pi} \sqrt{k^2+m^2} \frac{d}{dk} \theta
(k) + 
\Delta M_{[b+f]} = 0
$ (see \cite{rebhan}, eq (59) ). The expression in (\ref{k99}) again gives the accepted
result (\ref{ura1}). Note that if one assumed either 0 or 2 fermionic
zero modes,
the answer would be infinite. 

In \cite{we} the corresponding analysis for periodic boundary conditions was
found to yield exactly the same result and we  shall use those conditions
in what
follows.

\section{Supersymmetric sphalerons on a circle }

We turn now to a justification of the rules that one should take two
bosonic zero
modes and one fermionic zero mode into account when using mode
regularization for
$M^{(1)}_s$.

Consider the first sphaleron solution on a circle with large
circumference $2L$;
it describes a kink-antikink system  with periodic boundary conditions
\cite{manton}. The bosonic fluctuations around this background have been analyzed
in \cite{manton}, and we quote the result. For $L=\infty$ there are two zero
modes, but when $L$ is reduced (bringing the kink and the antikink 
together) one
of the zero modes becomes unstable ($\omega^2<0$), while rotational invariance
guarantees that the other zero mode remains at zero ($\omega^2=0$). The
value for
the unstable mode for large $L$ (adapted to our normalization)  is $
\omega^2 = - 48  m^2 \exp(-mL)
$.

We now extend the sphaleron solution to the case with fermions present.
We need
the background solution. It is given by
\be
\phi_n(x) = \frac{m k b}{\sqrt{2 \lambda}} {\rm sn}(bx,k)
\ee where $b = [1/(2(1+k^2))]^{1/2} m$ and ${\rm sn(bx,k)}$ is  an elliptic
function \cite{korn}. All we need is that this function satisfies the classical
field equations, is odd in $x$, where $x=0$ is the center of the kink (or
antikink, of course), and smooth on the circle.

We can now settle the issue of the fate of the fermionic zero mode in
one line:
the Dirac Hamiltonian due to (\ref{dirac1}) is manifestly self-adjoint, so that $\omega$ is
real and
thus
$\omega^2$ is nonnegative. Thus all one has to study is whether there
are any
zero modes, and how many. For $\omega=0$ the Dirac equation in (\ref{dirac4}) and (\ref{dirac4'}) has as
solutions 
\be
\label{tufta18}
\psi_I = \beta
\left(
\begin{array}{c} a_1 \exp [- 2 b k \int {\rm sn} (bx,k) dx ] \\  0
\end{array}
\right), 
\hspace{1cm }
\psi_{II} = \gamma 
\left(
\begin{array}{c} 0 \\ a_1 \exp [ 2 b k \int {\rm sn} (bx,k) dx ] 
\end{array}
\right) .
\ee Because the function ${\rm sn}(bx,k)$ is odd in $x$, the spinors at opposite
points from the center of the kink  are equal, and hence if one goes
around the
circle from $x=0$ in either direction, one reaches the same value for
the spinor
at the antipodal point. Thus the solutions are continuous. They are actually
smooth (differentiable) because the Dirac equation is first order in derivatives.

Our conclusion is that for a sphaleron background the fermionic spectrum
has two
zero modes, the same as for an infinitely separated kink and antikink. The
canonical equal-time anticommutation relations read according to the Dirac
formalism $\{
\psi_i(x,t), \psi_j(y,t) \} = \delta_{ij} \delta(x-y)$. It follows that the
operators $\beta$ and $\gamma$ satisfy $\beta^2=\gamma^2=1$ and $\{
\beta, \gamma
\}=0$: hence
$\beta+i
\gamma$ and $\beta - i \gamma$ form the annihilation and creation
operators for
one degree of freedom.

It is possible to give a physical explanation why for fermions 
$\omega^2 > 0$ for the $
\bar{K} K$ system but $\omega^2=0$ for the sphaleron system. Consider a
configuration $\phi(x)$ for the boson field on the circle vanishing at
two points
$x=x_1$ and $x=x_2$ which are not antipodal, such that $\phi(x)$ is a
solution of
the field equation except at $x=x_1$ and $x=x_2$. (Such a solution
exists because
it describes according to  the usual mechanical analogue a ball oscillating
around the bottom of the inverted potential.  For the segment where
$x_1$ and
$x_2$ are nearer to each other, the value of
$\frac{\partial}{\partial x} \phi (x)$ at $x_1$ and $x_2$ is smaller
than for the
other segment where $x_1$ and $x_2$ are further apart.) By making the
radius of
the circle large enough, the effect of the kink and antikink on the
fermions can
be neglected. (The kink and antikink fields only differ from their asymptotic
values over a distance $\Delta x \sim \mu^{-1}$). Note that the zero
mode of the
kink and antikink increases exponentially on one segment while it decreases
exponentially on the other segment. It becomes clear that a spinor which is
transported along the circle can not be periodic because the segments have
different length.

Thus in the sphaleron background the assumption $\omega=0$ can not be satisfied. For large $L$ but fixed
$x_1-x_2$ this looks like the $ \bar{K} K$ solution on the infinite line except
that now the discontinuities in $\phi^\prime(x)$ arise at the centers of
the kink
and antikink, instead of in between.

\section{Zero modes of the $\phi_{\bar{K} K}$ system} We now study the  discrete
spectrum of the kink-antikink system on an infinite line with the
background of
Fig. 1.

The bosonic modes are the solutions of (\ref{bof}). The solutions for
$x>0$ are given by \cite{raj}
\be
\label{uf88}
\eta_K(\omega,x)= \exp{  i k  (x-\frac{L}{2}) } \left[ -3 \tanh^2
\frac{m(x-\frac{L}{2} )}{2} +1+ 
\frac{4 k^2}{m^2} + \frac{6 i  k}{m} \tanh \frac{m(x-\frac{L}{2} )}{2} \right]
\ee  with $\omega^2=k^2+m^2$. For the solutions with $\omega^2<m^2$ on
$0 \le x <
\infty$ which are square-integrable, we take
$k = i
\kappa$ with $\kappa>0$.  The solution of ($\ref{bof}$) should also have
the same
left- derivative as right- derivative at $x=0$ because the potential is
continuous.

When one considers the zero modes of the kink and the antikink together,
taking the symmetric combination, the resulting function is still
continuous at
$x=0$, but the derivative is discontinuous (there is a cusp). Making
$\omega^2<0$ decreases the curvature of the solution, and one can find a
value of
$\omega^2$  such that also the derivative becomes continuous. Hence, the lowest
mode in the $\bar{K} {K}$ system is symmetric and has negative
$\omega^2$. Using (\ref{uf88})  one finds  $\omega^2=-12 m^2 \exp(-mL)$
for large
$L$. The next mode is antisymmetric. One can find the value of
$\omega^2$ for this
solution by requiring that $\psi_K(\omega,x)=0$ at $x=0$. There is no further
condition involving derivatives, because  if we take the antisymmetric  
combination of the solutions which vanishes at $x=0$, its derivative is there
continuous. We find that this second mode has $\omega^2>0$, namely $\omega^2=12
m^2
\exp(-mL)$ for large $L$. Note that contrary to the sphaleron there is
now no
longer a zero mode. The reason is clear: the zero mode would be the
derivative of
the configuration $\phi_{\bar{K} K}(x)$ in Fig. 1, but this derivative is
discontinuous at $x=0$, and therefore not a solution.

Let us now turn to the fermionic sector. The fermionic modes  which
should become
zero modes as $L$ tends to infinity are given by 
\be
\label{f27}
\psi(x,t) = \alpha \left( \begin{array}{c} \psi_K^R(\omega,x) \\ \frac{i}{\omega}
(\partial_x+c \sqrt{2 \lambda} \phi_K(x) ) \psi_K^R(\omega,x)  \end{array}
\right) e^{-i \omega t} +
\beta \left( \begin{array}{c} \psi_K^R(\omega,x) \\ -\frac{i}{\omega}
(\partial_x+c \sqrt{2 \lambda} \phi_K(x) ) \psi_K^R(\omega,x)  \end{array}
\right) e^{i \omega t} 
\ee  for positive $x$, and  
\be
\label{f28}
\psi(x,t) = \gamma \left( \begin{array}{c}  \frac{i}{\omega} (\partial_x-c
\sqrt{2 \lambda} \phi_{\bar{K}}(x) ) \psi_K^L(\omega,x) \\
\psi_K^L(\omega,x) 
\end{array}
\right) e^{-i \omega t} +
\delta \left( \begin{array}{c}  -\frac{i}{\omega} (\partial_x-c \sqrt{2 \lambda}
\phi_{\bar{K}}(x) ) \psi_K^L(\omega,x) \\  \psi_K^L(\omega,x)  \end{array}
\right) e^{i \omega t} 
\ee for negative $x$.  The function $\psi_K^R(\omega,x)$ is the solution
of the
Schr\"odinger equation (\ref{uf6}) with frequency $\omega$ which
vanishes for $x
\to \infty$. Similarly, $\psi_K^L(\omega,x) $ is the solution of (\ref{uf6})
which vanishes for $x \to -
\infty$. Further, 
$\phi_K(x) = m / \sqrt{2 \lambda} \tanh (x-L/2) $ for $x>0$ and
$\phi_{\bar{K}}(x) = - m / \sqrt{2 \lambda} \tanh (x+L/2) $ for
$x<0$.  For nonzero $\omega$ continuity at the origin fixes $\omega$
\be
\label{cnt}
\omega^2 \psi_K^L \psi_K^R = - \left[  (\partial_x +c \sqrt{2 \lambda}
\phi_K )
\psi_K^R  \right]  \left[  (\partial_x -c \sqrt{2 \lambda}
\phi_{\bar{K}} )
\psi_K^L  
\right] \ \ {\rm at } \  \ x=0
\ee  The expressions for $\psi_K^R$ and  $\psi_K^L$ are given by
\cite{raj} 
\be
\label{ex24u}
\begin{array}{c}
\psi_K^R(\omega,x)= \exp{ - \kappa  (x-\frac{L}{2}) } \left[ -3 \tanh^2
\frac{m(x-\frac{L}{2} )}{2} +1- 
\frac{4 \kappa^2}{m^2} - \frac{6  \kappa}{m} \tanh \frac{m(x-\frac{L}{2} )}{2}
\right] \hspace{1cm} {\rm for} \hspace{2mm} x>0
\\
\psi_K^L(\omega,x)= \exp{  \kappa  (x+\frac{L}{2}) } \left[ -3 \tanh^2
\frac{m(x+\frac{L}{2} )}{2} +1- 
\frac{4 \kappa^2}{m^2} + \frac{6  \kappa}{m} \tanh \frac{m(x+\frac{L}{2} )}{2}
\right] \hspace{1cm} {\rm for} \hspace{2mm} x<0
\end{array}
\ee  with $\kappa = (m^2-\omega^2)^{1/2}>0$. For large $L$ the frequency $\omega$
is small, and expanding (\ref{cnt}) in powers of $\omega^2$ one finds to leading
order
\be
\label{tufta24}
\omega^2 = \frac{9}{4}m^2  \frac{(1-\tanh \frac{mL}{4})^4 }{(2-\tanh
\frac{mL}{4})^2} \sim 36 m^2 \exp (-2 m L) 
\ee Note that $\omega^2$ tends to zero as $\exp(-2mL)$ for large $L$. As
$\omega^2$ is positive, the frequencies in (\ref{f27}) and  (\ref{f28})
are real. 

The results for $\omega^2 \approx 0$ for the bosons and fermions in a sphaleron
background display a suggestive relation to the same results for a
$\bar{K} K$ background. For the bosons one finds $\omega^2=-48 m^2
e^{-mL}$ and
$\omega^2=0$ in a sphaleron background, but $\pm 12 m^2 e^{-mL}$ in a
$\bar{K} K$ background. A tunneling argument shows that for the $\bar{K}
K$ system
the two zero mode levels become split symmetrically around zero, yielding
$\omega^2= \pm A e^{-mL}$ (with $A=12  m^2$). The factor $e^{-mL}$ can be
explained by considering the zero mode  of the kink in a potential
$V+\Delta V$
where
$V$ is the potential
 of the kink and $\Delta V$ is the potential due to the antikink. Perturbation
theory yields then for the diagonal correction to $H=\omega^2$ a value
$\int dx [ \eta \Delta V \eta ] \sim e^{-2mL}$. However, the main effect
is an
off-diagonal mixing, yielding $\omega^2=\int dx [\bar{\eta} \Delta V
\eta]$ where
$\bar{\eta}$ is the zero mode of the antikink. Diagonalization of the
$2 \times 2$ mixing matrix indeed produces $\omega^2 \sim \pm e^{-mL}$.
In fact,
we claim that the value of the splitting in $\omega^2$ in the sphaleron
case is
$-4A $. The reason is that the magnitude of the splitting in the sphaleron
background  should be twice the magnitude of the splitting in the
$\bar{K} {K}$  background because in the sphaleron case forces work in both
directions.
 Also the results for the fermionic distorted zero modes can be
explained. The
factor
$e^{-2mL}$ in $\omega^2$ given by (\ref{tufta24}) for the $\bar{K} K $ system is due to the overlap effect in 
$\Delta H = \omega \sim \int \psi_1 \Delta V \psi_2 dx$ yielding $\omega \sim
e^{-mL}$, (and thus $\omega^2 \sim e^{-2mL}$), as $\psi_1 \sim e^{-mL} $ where
$\Delta V$  is of order unity. This also explains why $\omega^2$ for the fermions
is positive. 

The continuity of (\ref{f27}) and (\ref{f28}) at the origin $x=0$
requires 
$\alpha = g
\gamma$ and $\beta = - g \delta$ where 
\be 
\label{uf25} g = \frac{i  \left[ \partial_x - c \sqrt{2 \lambda}
\phi_{\bar{K}}(0) \right]
\psi_K^L(\omega,0) }{ \omega \psi_K^R(\omega,0)}.
\ee  The expression (\ref{f27}) can then be rewritten as
$$
\label{f27f}
\psi(x,t) = g \left[  \gamma \left( \begin{array}{c} \psi_K^R(\omega,x) \\
\frac{i}{\omega} (\partial_x+c \sqrt{2 \lambda} \phi_K(x) )
\psi_K^R(\omega,x) 
\end{array}
\right) e^{-i \omega t}  \right.  \left. \right.
$$
\be 
\label{uf26} -
\left.
\delta \left( \begin{array}{c} \psi_K^R(\omega,x) \\ -\frac{i}{\omega}
(\partial_x+c \sqrt{2 \lambda} \phi_K(x) ) \psi_K^R(\omega,x)  \end{array}
\right) e^{i \omega t} \right] \ \ \ {\rm for} \ x \ge 0
\ee  while (\ref{f28}) is unchanged.  Reality of $\psi(x,t)$ implies
$\alpha=\beta^*$ and $\gamma= \delta^*$, while $g^*=-g$. This shows that we
really have only one set of creation and annihilation operators, $b=g
\gamma$ and
$b^*=-g \delta$. Notice also, that when the separation $L$ is large, the
expressions  
$ (\partial_x+c \sqrt{2 \lambda} \phi_K(x) ) \psi_K^R(\omega,x)$ and  
$(\partial_x-c \sqrt{2 \lambda} \phi_{\bar{K}}(x) ) \psi_K^L(\omega,x)$
are very
small( they vanish at $\omega^2=0$ because of the BPS equation, and are
of order
$\omega^2$ for nonzero $\omega$).  As a result, near $x=L/2$ both
solutions have
mostly an upper component and a negligible lower component, while near
$x=-L/2$ the opposite holds. However near $x=0$ the upper and lower
components of
both solutions are of equal magnitude (although much smaller then the leading
components at $x=\pm L/2$).

From the equal-time canonical anticommutation relations of $\psi(x,t)$
as given
by the Dirac formalism for Majorana spinors $\{ \psi_i(x,t),\psi_j(y,t)
\} =
\delta_{ij} \delta(x-y)$ one reads off the anticommutators for the
$\gamma$ and
$\delta$. One finds (after properly normalizing the wave functions for the
almost-zero modes) 
\be
\{ \gamma, \gamma \} =0, \ \ \{ \delta, \delta \} = 0, \ \ \{ \gamma,
\delta \} = 1 
\ee  Reality of $\psi (x,t)$ implies 
\be
\gamma = \delta^\dagger
\ee Shifting the origin in time from $t=0$ to $t=\tau$ leads to operators
$\gamma(\tau) = e^{- i \tau \omega} \gamma$  and $\delta ( \tau) = e^{i \tau
\omega} \delta$. Independence of $\tau$ implies $\{ \gamma(\tau), \delta(\tau)
\}=1$ and
$\{ \gamma ( \tau), \gamma ( \tau) \}= \{ \delta ( \tau) , \delta (
\tau) \} =0 $, which are obviously true. 

For $\omega \to 0$ ($L \to \infty$) one can introduce two hermitian operators
which commute with the Hamiltonian
\be
\begin{array}{c} b = ( e^{i\omega \tau} \gamma (\tau) + e^{-i
\omega \tau}
\delta(\tau) ); \ \
\ \ b=b^\dagger, \ \ b^2=1, \\ d = \frac{1}{i} ( e^{i\omega \tau}
\gamma(\tau) -
e^{-i\omega \tau}
\delta(\tau)); \ \ \ \ d=d^\dagger, \ \ d^2=1, \\
\end{array}
\ee  The operators $b$ and $d$ are then the operators for the zero modes
of the
kink and antikink, respectively.

\section{The isolated kink}

Having understood mode counting for the kink-antikink system on the
circle and on
the infinite line, we now turn to the problem of the kink alone. This section
consists of three parts: (i) it begins with an analysis of the three
contributions to the energy density (localized near the kink, localized
near the
boundary or uniformly distributed ), (ii) next we make explicit computations
which corroborate the general analysis and (iii) we interpret the
results in
terms of a Z$_2$ gauge symmetry.

\subsection{Localized and delocalized energy}

We shall begin with  a very simple approach which is guaranteed to yield the
correct answer for the kink mass
\cite{we}: given the solutions for the kink-antikink system with (for
definiteness) periodic boundary conditions at
$x = \pm L$, we just look at the behavior of these solutions halfway
between kink
and antikink, and this determines a  set of boundary conditions at $x=L$ and
$x=0$ for a kink centered at $x=\frac{L}{2}$.   If we use the corresponding
frequencies in the mass formula, we must get the correct result, namely
half the
mass shift for the
$\bar{K} K$ system.  These boundary conditions are periodic (P) or antiperiodic
(AP) for the boson field fluctuations in both sectors, and also P and AP for
fermions in the trivial sector.  For fermions in the the kink sector, the
conditions are twisted periodic (TP) or twisted antiperiodic (TAP)  [specifically,
$\psi_{1}(-L/2)=\psi_{2}(L/2)$ and $\psi_{2}(-L/2)=\psi_{1}(L/2)$ for TP,
$\psi_{1}(-L/2)=-\psi_{2}(L/2)$ and $\psi_{2}(-L/2)=-\psi_{1}(L/2)$ for TAP)
].\footnote{ Using $\psi_1= e^{ik(x+L/2)}$
on the far left, $\psi_1=e^{ik(x+L/2+\delta+\theta)}$ near $x=0$ and 
$\psi_1=A e^{i(k(x-L/2)+ 2 \delta +\theta)}$ on the far right (see figure
2)  one
finds from continuity at the origin $A=e^{ikL}$. The Dirac equation
(\ref{dirac1}) yields $\psi_2$, in particular $\psi_2(L)=
e^{i(3kL/2+2\delta+3\theta/2)}$.  Imposing $\psi_1(-L)= \psi_1(L)$
(which implies
that also the derivative is periodic), one finds the quantization
condition 
$e^{i(2kL+2\delta+\theta)}=1$. Clearly $\psi_2(L)/\psi_1(0) =
e^{i(kL+\delta+\theta/2)}=\pm 1$. }  Evidently, taking both P and AP
conditions (or TP and TAP conditions) would overcount the number of states in the interval $0 \le x
\le L$ by a factor two, so one may take the contributions from each and then
average the results.  

In particular, by identifying the kink mass as half the $\bar{K}K$ mass, one
should take for the fermions the  difference between the averages of
mode sums
with P and AP boundary conditions in the trivial sector, and sums with
TP and TAP
boundary conditions in the kink sector.   There is no fermionic P or AP solution
with $\omega^2 \sim 0$ in the trivial sector, and clearly therefore no
corresponding fermion degree of freedom.  (The general solution of the Schr\"odinger equation $\psi_1=\alpha
e^{\kappa x}+ \beta e^{- \kappa x}$ with $\kappa^2=m^2-\omega^2 \sim
m^2$ can not
be P or AP for both $\psi_1$ and $\psi_2$). As there is only one fermionic
zero-frequency TP or TAP {\bf solution}  in the kink sector (see below),
there are
no zero-frequency fermion {\bf degrees of freedom} in either sector .  However,
in the kink-antikink system, there was one fermionic degree of freedom
near zero
frequency, see (\ref{uf26}), hence the highest-energy mode in the kink sector
must be counted with half the weight of any other mode.  This may be
accomplished, for example, by omitting the highest energy state in the
sum for
twisted antiperiodic boundary conditions\footnote{  For future use we do
at this
moment a little calculation. Having $N$ modes for TP conditions in the kink
sector and $N-1$ modes for TAP conditions in the kink sector, one finds
that the
difference of the mode sums is finite but nonzero. For $N=2M-1$ one finds
$$
\frac{1}{2} \sum \omega_{TP} - \frac{1}{2} \sum \omega_{TAP}
= \frac{1}{2} \omega_{7),N} +
 \sum_{n=1}^{M-1} \omega_{7)} -  \sum_{n=1}^{M-1}
\omega_{8)}=
\frac{\sqrt{\Lambda^2+m^2}}{2}+ \int_0^\Lambda \frac{dk}{2 \pi} (-\pi)
\omega^\prime =  
\frac{m}{2}
$$ (where we put $\omega_{7)}$  ($\omega_{8)}$) for the frequency with $k$,
satisfying TP (TAP) conditions in the kink sector, which we define in
our main
text above equation (\ref{po1}) ). Our arguments in the main text lead to the conclusion
that one
needs the average of the TP and TAP  conditions, but it is clear from
this little
calculation that using only TP or TAP conditions yields an incorrect finite
answer for
$M^{(1)}$. Obviously, taking instead $N-1$ modes with TP conditions and
$N$ modes with TAP conditions leads to a divergent value for the
difference (but
the same result for the average). }.  Clearly this prescription violates the
principle of summing equal numbers of modes, which is the basis of mode
regularization.

The contradiction is resolved by noting that the mode regularization
principle is
based on fixed boundary conditions, so that one may think of the
individual modes
as `particle wave functions' which are deformed in their
$x$-dependence and shifted in frequency as the background classical
field changes
from trivial to kink.  However, in the approach described above, the boundary
conditions also change between the two sectors   (from P and AP to TP
and TAP ).
As noted by Goldstone and Wilczek for complex (Dirac) fermion fields, the
180$^{\rm o}$ chiral rotation in (\ref{uf30}) of the Yukawa coupling
leads to a
flow of 1/2 unit of fermion charge out of the region \cite{gw} (see figure 4). 

$$
\epsfbox{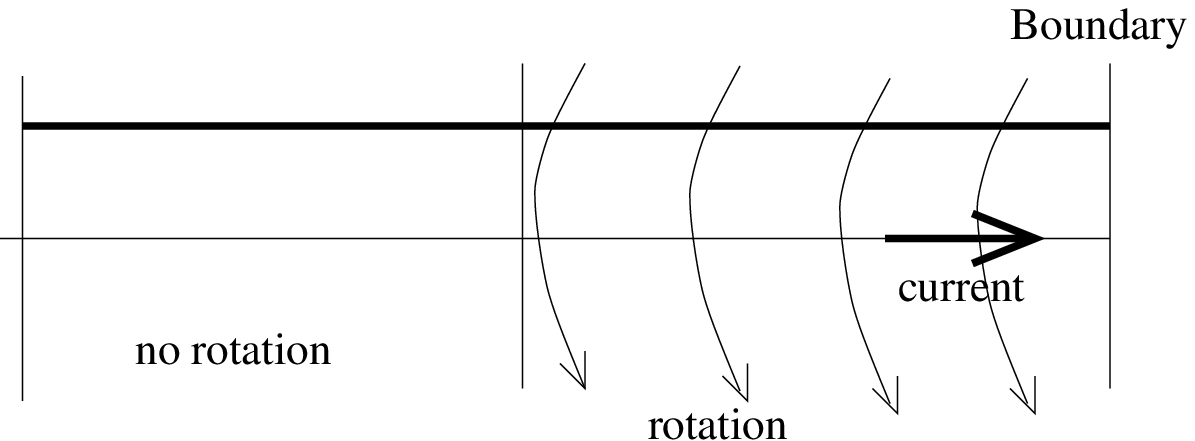}  $$
Figure 4. {\it The chiral rotation causes a current which leads to the accumulation of half a degree of freedom at the boundary
if the boundary conditions are kept fixed.
} \\ \\

In this
way they
gave a dynamical mechanism for the phenomenon which had been
discovered 
by Jackiw and Rebbi
\cite{jackiw}, that a kink coupled to Dirac fermions carries half-integer
fermionic charge. Fixed boundary conditions would stop this flow at the boundary,
but it is obvious that simultaneously rotating the boundary conditions would
maintain the flow.  This certainly is consistent with the actual result
in our
case, that (on averaging between P and AP in the trivial sector, and TP
and TAP
in the kink sector) half a Majorana fermion mode must be omitted from
the sum in
the kink sector. Drawing a superficial analogy (which will become less
superficial as we go on): having a complex Dirac fermion allows the same chiral
rotation (and loss of half a unit of fermionic charge) as having two
sets of
boundary conditions for a real fermion (where the chiral rotation leads
to the
loss of half a degree of freedom). 

The recipe obtained from the $\bar{K}K$ system gives a reasonable interpretation
for the kink alone, but does not satisfy the principle of mode
regularization that
one should use fixed boundary conditions. If we wish to  use fixed boundary
conditions, we run into the following problem:  The periodic conditions
in the
trivial sector and twisted periodic conditions in the kink sector both are
associated with `locally invisible' boundaries (namely, with {\bf plane wave}
solutions).   That is, they give no structure associated with the precise
location of the boundary \cite{misha}.  This means that, aside from the energy
density localized around the kink, in principle the only other possible
contribution would be a translationally invariant piece, corresponding
to an
energy density
${\cal O} (1/L)$ as we shall explain.
$$
\epsfbox{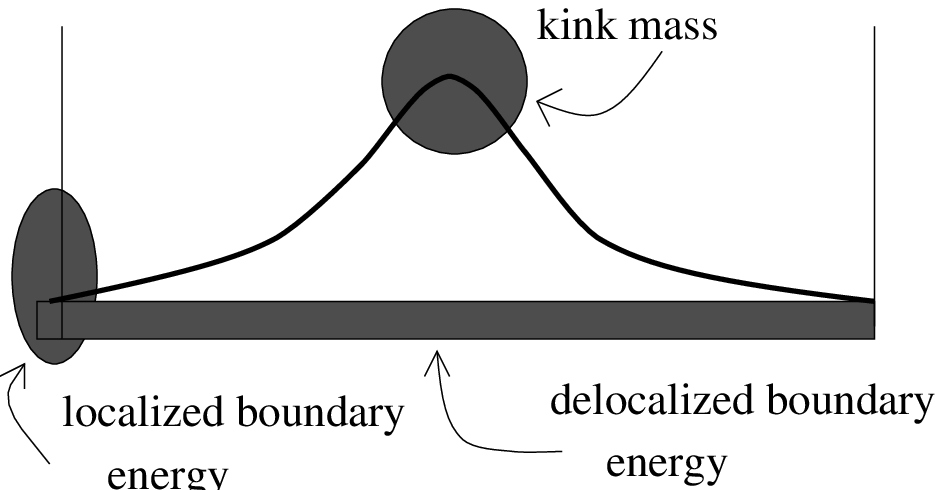}  $$
Figure 5. {\it The total energy density in the kink background.} \\ \\
Except for this possibility, 
computing the
quantum correction using these boundary conditions is identical in
effect to the
procedure advocated by Shifman et al. \cite{shifman}, to compute the
local energy
density and integrate it over the region of the kink.   However, either periodic
conditions in the kink sector or twisted periodic conditions in the trivial
sector would produce a `visible' boundary, forcing {\bf standing wave} rather
than plane wave solutions.  
In the fermionic sum $\sum \omega_n^f$ there is then a true contribution to the mass and a boundary 
contribution ${\rm E_{\rm boundary}}$.
Therefore, one obtains an energy shift 
$M^{(1)}_{\rm kink} - {\rm E}_{\rm boundary}$ if one uses fixed (i.e., in both
sectors) P plus AP boundary conditions,  and 
$M^{(1)}_{\rm kink} + {\rm E}_{\rm boundary}$ with fixed TP (plus TAP) boundary
conditions. By averaging the two forms one obtains
$\Delta M_{\rm kink}$ by itself.

To spell this out further: P or AP conditions in the trivial sector have no
localized boundary energy $E_{\rm boundary}$. TP or TAP conditions in
the kink
sector also have no localized $E_{\rm boundary}$. However, TP and TAP in the
trivial  sector, as well as P and AP in the kink sector, all have localized
boundary energy
$E_{\rm boundary}$. Making a chiral rotation near the boundary which
maps $\phi
\to -\phi $ and twists the fermions 
\be
\label{uf30}
\left( \begin{array}{c}
\psi_1 \\ \psi_2 \\ \end{array} \right)^\prime = e^{\pm i \frac{\pi}{2} \sigma_1}
\left( \begin{array}{c}
\psi_1 \\ \psi_2 \\ \end{array} \right) = \pm i \left( \begin{array}{c}
\psi_2 \\ \psi_1 \\ \end{array} \right), \hspace{1cm} \phi^\prime =
e^{\pm i \pi
} \phi
\ee the localized boundary energy should not change\footnote{ As the
fermions are
Majorana,  one really should first complexify them (going to an $N=2$
model), but
one  can achieve the same goal by summing over both chiral rotations.}.  (In
addition, to keep the fermions real one needs a finite local gauge transformation
$\psi_1
\to i \psi_1$ and
$\psi_2 \to i \psi_2$.) Thus P $\to$ (TP,TAP) and AP $\to$ (TAP,TP) and
vice-versa.

Accepting that we need fixed boundary conditions (i.e. the same in the trivial
and in the kink sector) for mode regularization, and observing that we need
different boundary conditions to cancel the localized boundary energy,
we are
compelled to average over all four sets (P, AP, TP and TAP in both
sectors) of
boundary conditions for the fermions. The need for all four sets of boundary
conditions to implement the chiral symmetry is similar to the need to
consider all
four spin structures
 for the string on a torus. The conclusion is that boundary energy
occurs in the
various cases as indicated in table I.
\begin{center}
\begin{tabular}{|c||c|c|}
\hline  &  \multicolumn{2}{|c|}{Boundary conditions} \\ 
\cline{2-3}  Sector   &  P and AP & TP and TAP \\ 
\hline \hline  Trivial & 0 & ${\rm E_{\rm boundary} } $ \\   \hline 
Kink & ${\rm
E_{\rm boundary} } $  & 0 \\
\hline
\end{tabular}
\vspace{0.3cm}  \\ Table I. {\it Localized boundary energy for different sectors
and boundary conditions.}
\vspace{0.2cm} 
\end{center} Obviously, the boundary energy cancels in the average over
all four
choices with fixed boundary conditions.

What about the delocalized boundary energy? In the trivial sector  with
P and AP
conditions, the difference of the P sum of $\sqrt{\left( \frac{2 \pi
n}{L}\right)^2 + m^2 }$ and the AP sum  of $\sqrt{\left( \frac{(2n+1)
\pi }{L}
\right)^2 + m^2 }$ can be grouped into a sum over quartets of states, starting
from the bottom. In each quartet the leading nonvanishing term is of order
$1/L^2$, but summing over all modes, the total energy difference is of order
$1/L$\footnote{ Using $\sqrt{\left( \frac{\pi (n+1)}{L} \right)^2 +m^2}- 2
\sqrt{\left( \frac{\pi n}{L}
\right)^2 +m^2}+ \sqrt{\left( 
\frac{\pi (n-1)}{L} \right)^2 +m^2} = \left( \frac{\pi}{L} \right)^2 \frac{d^2
\omega}{dk^2}+...$ with $n=2k+1$, one finds that the total energy
difference is
equal to $\frac{\pi}{2 L}$. } .  Hence, one can forget in the
computation of the
kink mass the delocalized boundary energy in the trivial sector.  This
permits us
to choose the average of the P sum and the AP sum as the energy of the trivial
vacuum, which we define to be zero.  In the kink sector one has locally invisible
boundaries for TP and TAP conditions,  so in these cases there could
also be
delocalized boundary energy. From the explicit calculation that TP (with $N$
modes)  minus TAP (with $N-1$ modes ) gives $m/2$, we conclude that
there are
different amounts of delocalized boundary energy  in the TP and TAP
sectors, so
delocalized energy does, in fact, occur\footnote{ We assume here that after
subtracting the localized and the delocalized boundary energy, one
obtains the
true mass of the kink which should not depend on the boundary conditions
\cite{shifman}. Since in the TP and TAP  sectors of the kink there is no
localised boundary energy, the difference $m/2$ must be due to delocalized
boundary energy. }.  Nevertheless, returning to our $\bar{K} K$ system,
we notice
that no delocalized energy could be created when one locally pulls the trivial
configuration
$ \phi =
\mu / \sqrt{\lambda } $ down  to the nontrivial configuration with a $\bar{K}K$,
and then separates $K$ and $\bar{K}$.  Consequently, when the average is taken,
the delocalized boundary energy contributions  must cancel\footnote{ These
statements follow from a standard assumption in field theory, that in the
presence of a mass gap, all correlators of observables fall
exponentially with
separation of arguments. Consider any correlator involving some number of
factors  $\left( \phi^2(x) -
\frac{\mu^2}{\lambda} \right) $ and one factor $\epsilon(x)$, the local  energy
density. By the general principle $\langle \epsilon (x) \rangle$ must fall
exponentially for $x$ far from kink and from antikink. Thus, there can
be no
translationally invariant piece of the energy density. From our analysis of the kink-antikink system we know that
the average over TP
 and TAP in the kink sector and   P and AP in the trivial sector will
not produce
such a contribution. The novelty here is that for TP or TAP separately
there is a
finite difference, which must be attributed to a translationally
invariant energy
density. This fact suggests that some principle must require averaging
over both
TP and TAP corrections, excluding a delocalized energy of order m.  We
shall see
shortly that there is indeed such a principle}. 
We summarize the results on delocalized boundary energy in table II.
\begin{center}
\begin{tabular}{|c||c|c|c|c|}
\hline &  \multicolumn{4}{|c|}{Boundary conditions} \\ \cline{2-5}
Sector   &  P
& AP & TP & TAP \\ \hline \hline Trivial & 0 & 0 & 0  & 0 \\ \hline Kink
& 0 & 0 & 
$\frac{\hbar m}{4}$ &   $-\frac{\hbar m}{4}$ \\
\hline
\end{tabular}
\vspace{0.3cm}  \\ Table II. {\it Delocalized boundary energy for different
sectors and boundary conditions.}
\vspace{0.2cm} 
\end{center} 

What about the zero modes, and the correct counting of states in the
sums?  For
fixed TP conditions, one has a single zero-frequency solution attached
to the
boundary in the trivial sector, namely (\ref{mody1}) with $a_1 e^{-mL/2}=a_2
e^{mL/2}$, and a single one attached to the kink in the kink sector, namely
(\ref{mody0}) with $ a_1 /\cosh^{2}
\frac{mL}{2} =  a_2 \cosh^2 \frac{-mL}{2} $.   Thus,  no matter how one might weigh the contribution
of such a
single solution, the effect cancels exactly in the subtraction.  For
fixed P
conditions, there are no zero modes in the trivial sector, and two in
the kink
sector, namely one attached to the kink and one attached to the
boundary, given
by (\ref{mody0}) with $a_1$ and $a_2$ arbitrary.   For TAP and AP
conditions the
same results hold\footnote{ Actually, to obtain two solutions with AP conditions
in the kink sector one needs exponentially small but nonvanishing
$\omega$. One may start with the Schr\"odinger equation for $\psi_1$ and raise
$\omega^2$ such that $\psi_1$ vanishes at $x=\pm L/2$. Then the Dirac equation
yields two solutions for $\psi_2$ corresponding to $\pm
\omega$, which are antisymmetric. }. Thus one must omit one (above-threshold) fermion mode from the Casimir 
sums over P and AP boundary conditions. Previously, from our study of the kink-antikink system, we were led to consider only TP and TAP conditions and then 
we needed to omit one term from their sum. Now we have a different message: we consider all four sets of
P, AP, TP and TAP, and omit one term from the P sum and one term from the AP sum.

Averaging over the four cases of fixed nontwisted
and twisted
boundary conditions once again leads to a reduction in the kink sum for the
fermions by half a fermion mode, but now the accounting is completely
straightforward, unambiguous and canonically justified. 

We have reached the following conclusions:  
\\ (i) for fixed boundary conditions no change in the total number of
degrees of
freedom occurs, as expected,  
\\ (ii) the issue how to weigh a single zero-frequency  solution in the
sums over
modes need not be solved (although we shall solve it) because both the trivial
and the kink sector have one such solution for twisted boundary
conditions, 
\\ (iii) In going from P and AP in the trivial sector to TP and TAP in
the kink
sector, $1/2$ degree of freedom is lost on the average\footnote{
To avoid confusion, note that for the calculation of the kink mass 
one must take equal numbers of modes in the trivial and kink sector for each 
set of boundary conditions, but what these numbers are is not important, they may differ from one set to another. 
} (it is radiated
away at
the boundary, in agreement with \cite{gw}. When the boundary conditions change
there is no reason to  expect that the number of degrees of freedom
remains the
same). 
\\ (iv) One must average over the four sets of boundary conditions for the
following reasons (a) in order to get the correct answer (the same as
from the
$\bar{K}K$ system), (b) the contributions from TP or TAP in the kink
sector are
different, so one may expect to need a particular combination of both, (c)
the chiral rotation from one set of boundary conditions in the trivial
sector 
links to two sets of boundary conditions in the kink sector, (d) in order that the localized boundary energy
of table I cancels, and (e) in order that the delocalized boundary 
energy given in table II cancels. 
\\ (v) For fixed boundary conditions, no degrees of freedom are lost,
but now one
zero-frequency solution can be attached to the boundary or to the kink. More
specifically, for visible boundary conditions there is always one zero-frequency
solution attached to the boundary, and when a kink is present there is
always one
zero-frequency solution attached to the
 kink. This yields four possible contributions, see table III.  

\begin{center}
\begin{tabular}{|c||c|c|c|c|}
\hline &  \multicolumn{4}{|c|}{Boundary conditions} \\ \cline{2-5}
Sector   &  P
& AP & TP & TAP \\ \hline \hline Trivial & 0 & 0 & 1 & 1 \\ \hline Kink &
2 & 2 & 1
& 1 \\
\hline
\end{tabular}
\vspace{0.3cm} \\  Table III. {\it Number of $\omega \approx 0$
solutions for
different sectors and boundary conditions.}
\vspace{0.2cm} 
\end{center}

\subsection{Explicit computations}

We now give some details. We start with periodic boundary conditions for the
bosonic fluctuations in the $\bar{K} K$ system:
$\eta(-L)=\eta(L)$ and $\eta^\prime(-L)=\eta^\prime (L)$. We could use the sphaleron solution discussed before as
background, or we could use
$\phi_{\bar{K}K}(x)$ to avoid elliptic functions at the price of not
having a
solution of the classical field equations at $x=0$.  For the
fluctuations this
makes negligible (exponentially small) difference. 

The quantization condition for the double system is $ 2kL + 2 \delta(k)
= 2 \pi
n$, $-\infty <n< \infty$. For the single kink we find two sets of boundary
conditions: P and A
\be
\begin{array}{c}
\eta(0) = \eta(L) \ \ {\rm and} \ \ \eta^\prime(0) = \eta^\prime(L) \ \
\ \ {\rm (P)} \\ 
\eta(0) = - \eta(L) \ \ {\rm and} \ \ \eta^\prime(0) = - \eta^\prime(L)
\ \ \ \ {\rm (A)} 
\end{array}
\ee With the solution $\eta(x,t) \sim \exp(i[kx \pm \frac{1}{2}
\delta(k)])$ we then find the quantization conditions
\be
\begin{array}{c} kL + \delta(k) = 2 n \pi \ \ \  {\rm (P)} \\ kL +
\delta(k) = (2 n +1) \pi \ \ \  {\rm (A)}
\end{array}
\label{ee32}
\ee where in both cases $-\infty <n< \infty$. Clearly we find for a
single kink
the same set of momenta  in (\ref{ee32}) as for the $\bar{K}K$ system.
The mass
formula for a single kink then reads
$$ M^{(1)}_{b} = \frac{1}{4}  \hbar \left[ 0 + \omega_B + m + 2 \sum_{n=2}^N
\omega_n^{P} - \sum_{n=-N}^N \omega_n^{P,(0)} \right] 
$$
\be
\label{bocont}
 +\frac{1}{4}  \hbar \left[ 0 + \omega_B + \sum_{n=-N}^{-2} \omega_n^{A}
+\sum_{n=1}^{N-1} \omega_n^{A}   - 2 \sum_{n=0}^{N-1} \omega_n^{A,(0)}
\right] + \Delta M_{b}
\ee  
$$ = \frac{\hbar \omega_B}{2} - \hbar m - \hbar \int_0^\Lambda
\frac{dk}{2 \pi}
\omega^\prime
\delta + \Delta M_{b}
$$ We used that in the nontrivial periodic case the solution with $n=0$ is
excluded, and the solutions with $n=+1$ and $n=-1$ both yield $k=0$,
giving the
term $m$. In the nontrivial antiperiodic case the solutions  with $n=-1$ and
$n=0$ are excluded. This yields (by construction) the correct mass for the
bosonic kink. (In fact the P and AP conditions each give the same
correct value
for the bosonic fluctuations).  In the $\bar{K}K$ system we had two
zero modes,
and this corresponds  to having one zero mode in each of the kink sectors.

Let us now turn to the fermionic fluctuations.  In this case the issue
of what
set of boundary conditions to use is much more subtle. As explained
earlier,  the
correct set of boundary conditions consists of periodic and
antiperiodic, both
twisted and untwisted, considered in both trivial and kink sectors. The formula
for the mass correction from the fermions then reads
\be
\label{princip}
\begin{array}{c} +  \frac{1}{4} (P+ A)_{Trivial \ sector} \\ -
\frac{1}{4} (P + A)_{Kink \ sector} \\ +  \frac{1}{4} (TP+ TAP)_{Trivial \
sector} \\ - \frac{1}{4} (TP + TAP)_{Kink \ sector} \\
\end{array} 
\ee 

As mentioned earlier, the first and the last lines appear naturally when one
reduces the  kink-antikink system to a single kink, so they do
separately give
the right result.  Now we check that the full set of sums also gives the correct
answer.

First, we address the issue of fermionic $\omega=0$ solutions under these
boundary conditions. The formal solutions of the Dirac equation with 
$\omega=0$, (\ref{dirac4}) and (\ref{uf9}), are given by (\ref{mody0})
for the
case of the kink background and by
\be
\label{mody1}
\psi_I = \left(
\begin{array}{c}
 {a_1}{e^{-mx}} \\  
 0
\end{array}
\right) , 
\hspace{3cm}
\psi_{II} = 
\left( 
\begin{array}{c} 0 \\ a_2 e^{mx} 
\end{array}
\right) 
\ee for the case of the trivial background 
$\phi_{triv}=\frac{m}{\sqrt{2 \lambda}}$. Adjusting the coefficients
$a_1$ and
$a_2$, one may try to satisfy the boundary conditions in
(\ref{princip}). This
results in the numbers of $\omega \approx 0$ solutions in particular sectors
given by
table III. 
 
Next we consider the continuous spectrum. The quantization conditions
for P and
AP boundary conditions in the trivial sector are obvious. 

We now address the TP  conditions in the trivial sector.  If one puts
\be
\label{fla1}
\psi_1 = e^{ikx}+ a e^{-ikx}
\ee  then it follows from the Dirac equation (\ref{dirac1}) that
\be
\label{fla2}
\psi_2=e^{i(kx+\frac{\theta}{2})} - a e^{-i(kx+\frac{\theta}{2})}
\ee 
where we define $\theta $ such that $e^{i
\frac{\theta}{2}}=-\frac{k}{\omega}+ i
\frac{m}{\omega}$. Twisted periodic conditions read
$\psi_1(0)=\psi_2(L)$ and 
$\psi_2(0)=\psi_1(L)$. We plug in $\psi_{1,2}$ and solve for $a$; this
gives the
following quantization condition
\be
\label{a38}
\frac{e^{i(kL+\frac{\theta}{2})}-1}{e^{-i(kL+\frac{\theta}{2})}+1}=a=\frac{e^{i
\frac{\theta}{2}}- e^{ikL}}{e^{-ikL}+e^{-i\frac{\theta}{2} }}
\ee 
which can be rewritten as $\sin kL = 0$, i.e. $kL=\pi n$. Notice
that if one
changes $k \to -k$ in (\ref{fla1}) and (\ref{fla2}), then
$e^{i \frac{\theta}{2}} \to -e^{-i\frac{\theta}{2} }$, so that $a \to
\frac{1}{a}$, and (\ref{fla1}) and (\ref{fla2}) stay the same up to
normalization.  Therefore negative
$k$ do not produce new independent solutions.  There is formally a
solution with
$n=0$, i.e. $k=0$, but for this solution (\ref{a38}) yields $a=-1$ and then (\ref{fla1}) and (\ref{fla2})
yield 
$\psi_1=\psi_2=0$ everywhere, so we must also exclude $n=0$.

For TAP conditions in the trivial sector one gets the same result (which is
obvious: one changes the sign of $\psi_2$ and then requires twisted periodic
conditions on
$\psi_1$ and $-\psi_2$).

The P boundary conditions in the kink sector correspond to standing wave
solutions of the Dirac equation (\ref{dirac1}) which were found in \cite{rebhan}.
The general solution of the Dirac equation for $\psi_1$ reads
$$ 
\psi_1(x)= \exp{  i k  (x-\frac{L}{2}) } \left[ -3 \tanh^2
\frac{m(x-\frac{L}{2} )}{2} +1+ 
\frac{4 k^2}{m^2} + \frac{6 i  k}{m} \tanh \frac{m(x-\frac{L}{2} )}{2} \right]
$$
\be + a 
\exp{  -i k  (x-\frac{L}{2}) } \left[ -3 \tanh^2
\frac{m(x-\frac{L}{2} )}{2} +1+ 
\frac{4 k^2}{m^2} - \frac{6 i  k}{m} \tanh \frac{m(x-\frac{L}{2} )}{2} \right]
\ee which gives  the following  asymptotic expressions for fermion components
\be
\psi_1= \left\{ 
\begin{array}{c} - e^{i(kx-\frac{\delta}{2})} - a e^{-i(kx-\frac{\delta}{2})},
\hspace{1cm}x \approx 0 \\ - e^{i(kx+\frac{\delta}{2})} - a
e^{-i(kx+\frac{\delta}{2})}, \hspace{1cm}x \approx L \\
\end{array}
\right.
\ee and, from the Dirac equation (\ref{dirac1}),
\be
\psi_2= \left\{ 
\begin{array}{c} e^{i(kx-\frac{\delta}{2}-\frac{\theta}{2})} - a
e^{-i(kx-\frac{\delta}{2}-\frac{\theta}{2})},
\hspace{1cm}x \approx 0 \\ e^{i(kx+\frac{\delta}{2}+\frac{\theta}{2})} - a
e^{-i(kx+\frac{\delta}{2}+\frac{\theta}{2})},
\hspace{1cm}x \approx L \\
\end{array}
\right.
\ee 
Requiring periodicity, it is clear\footnote{
Alternatively, an explicit calculation similar to (\ref{a38}) yields $\sin \left( \frac{kL+\delta }{2} \right)=0$ or
$\sin \left( \frac{kL+\delta+\theta }{2} \right)=0$.
} that it can be achieved either for
$a=1$, $kL+\delta+\theta=2 \pi n$ or for $a=-1$, $kL+\delta=2 \pi n$,
and only
positive $n$ are needed to produce distinct solutions. In particular for
$n=1$, $k=0$ one computes from the latter set $\psi_1|_{k=0}=\left[1-3 \tanh^2
\frac{m(x-L/2)}{2} \right] (1+a) = 0$, so by the Dirac equation
$\psi_2=0$. Thus this solution must be excluded, just like the $k=0$
solution in
TP of the kink sector.

The situation with AP conditions in the kink sector is quite analogous. The
antiperiodicity is achieved by putting $a=-1$, $ kL+\delta+\theta=2 \pi
n+ \pi$ or
by 
$a=1$, $ kL+\delta=2 \pi n+ \pi$. In particular there seems to be a
solution at
$n=0$, $k=0$ in the first set. This solution is excluded by the same
argument as
before: it is easy to check that in this case $\psi_1=\psi_2=0$ everywhere.

The TP conditions in the kink sector were worked out in
\cite{misha}; the computation for the TAP case in the kink sector is again
straightforward (it turns out that these two sets of boundary conditions
are the
only ones consistent  with plane wave solutions, see \cite{misha}).  

Thus, the quantization conditions for fermions are:
\vspace{0.2cm}
\\ I. {\bf Nontwisted}
\vspace{0.1cm}
\\1) Periodic trivial sector: $kL=2 \pi n$, all $n$.
\vspace{0.0cm}
\\2) Antiperiodic trivial sector: $kL=2 \pi n+\pi$, all $n$.
\vspace{0.0cm}
\\3) Periodic kink sector: a) $kL+\delta+\theta=2 \pi n$,  $n=1,2,3,...$ and
 b) $kL+\delta=2 \pi n$, $n=2,...$ .
\vspace{0.0cm}
\\4) Antiperiodic kink sector:  a) $kL+\delta+\theta=2 \pi n+\pi$,  $n=1,2,...$
and
 b) $kL+\delta=2 \pi n+\pi$, $n=1,2,...$ .
\vspace{0.0cm}\\ II. {\bf Twisted}
\vspace{0.1cm}
\\5) Twisted periodic trivial sector: a) $kL=2 \pi n$, $n=1,2,...$ and
 b) $kL=2 \pi n + \pi$, $n=0,1,2,...$.
\\6) Twisted antiperiodic trivial sector: a) $kL=2 \pi n$, $n=1,2,...$ and
 b) $kL=2 \pi n + \pi$, $n=0,1,2,...$.
\\7) Twisted periodic kink sector: $kL+\delta+\frac{\theta}{2}= 2 \pi
n$,  all
$n$, $n \ne 0$
\\8)  Twisted antiperiodic kink sector: $kL+\delta+\frac{\theta}{2}= 2 \pi
n+\pi$, all $n$, $n \ne 0, -1$ 
\vspace{0.2cm}
\\

We now work out the mass corrections due to fermions in the kink
background  for
each fixed set of boundary conditions separately. In all the sums we
keep equal
numbers of modes in both trivial and nontrivial sectors. (For these fermionic
corrections  we subtract the contributions in the kink sector  from the 
contributions in the trivial sector.)

We start with P boundary conditions. The fermions give the contribution
$$ M^{(1)}_f(P)=  \frac{\hbar}{2} \sum_{n=-N}^{N} \omega_{1)} -
 \frac{\hbar}{2} \sum_{n=1}^{N} \omega_{3a)} - \frac{\hbar}{2} 
\sum_{n=2}^{N} \omega_{3b)} - 0 - \frac{\hbar \omega_B}{2} + \Delta M_{f}
$$ \be =- \frac{\hbar \omega_B}{2}+{\hbar m } + \hbar \int_0^{\Lambda}
\frac{dk}{2 \pi}
\omega^\prime \left( \delta + \frac{\theta}{2} \right) + \Delta M_{f}
\label{po1} 
\ee  where we have taken into account one $\omega = 0$ fermionic degree of
freedom in the kink sector, see table III.  For AP boundary conditions
one obtains
$$ M^{(1)}_f(AP)= {\hbar} \sum_{n=0}^{N} \omega_{2)} -
 \frac{\hbar}{2} \sum_{n=1}^{N} \omega_{4a)} - \frac{\hbar}{2} \sum_{n=1}^{N}
\omega_{4b)} - 0 -
\frac{\hbar \omega_B}{2} + 
\Delta M_{f}
$$ \be = - \frac{\hbar \omega_B}{2}+{\hbar m } + \hbar \int_0^{\Lambda}
\frac{dk}{2 \pi}
\omega^\prime \left( \delta + \frac{\theta}{2} \right) + \Delta M_{f}
\label{po2} 
\ee  and again, one $\omega \approx 0$ fermionic degree of freedom in
the kink
sector is counted in the sum.  The result is exactly the same as in the
case of P
boundary conditions. 

For TP one does not have an $\omega \approx 0$ degree of freedom, so 
 $$ M^{(1)}_f({TP})= \frac{\hbar}{2} \sum_{n=1}^{N} \omega_{5a)} +
 \frac{\hbar}{2} \sum_{n=0}^{N} \omega_{5b)} - {\hbar} \sum_{n=1}^{N}
\omega_{7)}  - \frac{\hbar
\omega_B}{2} + \Delta M_{f}
$$ \be =  \frac{\hbar \sqrt{\Lambda^2+m^2}}{4}- \frac{\hbar \omega_B}{2}+\frac{
\hbar m }{4} + \hbar
\int_0^{\Lambda} \frac{dk}{2 \pi}
\omega^\prime \left( \delta + \frac{\theta}{2} \right) + \Delta M_{f}
\label{po3} 
\ee  For TAP, however, we get a result which is different from the result
(\ref{po3}) for TP:
$$ M^{(1)}_f(TAP)= \frac{\hbar}{2} \sum_{n=1}^{N} \omega_{6a)} +
 \frac{\hbar}{2} \sum_{n=0}^{N} \omega_{6b)} - {\hbar} \sum_{n=1}^{N}
\omega_{8)}  - \frac{\hbar
\omega_B}{2}  + \Delta M_{f}
$$ \be = - \frac{ \hbar \sqrt{\Lambda^2+m^2}}{4}- \frac{\hbar
\omega_B}{2}+\frac{3 \hbar m }{4} + \hbar
\int_0^{\Lambda} \frac{dk}{2 \pi}
\omega^\prime \left( \delta + \frac{\theta}{2} \right) + \Delta M_{f}
\label{po4} 
\ee 

Actually, only the averages ${\rm (P+AP)}/2$ and  ${\rm (TP+TAP)}/2$ are
invariant under the 
Z$_2$-gauge symmetry $\psi \to - \psi$. It is easy to compute these averages,
which of course do not have linear divergences. In terms of
$M^{(1)}_f$,  the fermionic contribution to the mass of the kink in (\ref{ura1}),
they can be written as
\be
\frac{1}{2} \left[ M^{(1)}_f(P) + M^{(1)}_f({AP}) \right] = M^{(1)}_f+
\frac{\hbar m}{4}  \label{po5}
\ee 
\be
\frac{1}{2} \left[ M^{(1)}_f(TP) + M^{(1)}_f(TAP) \right] =  M^{(1)}_f-
\frac{\hbar m}{4}  \label{po6}
\ee Note the presence of a `half-mode' contribution $\frac{\hbar m}{4}$
representing the boundary energy, which as expected appears with
opposite signs
in the (P+AP) and (TP+TAP) sums.  To find out whether this boundary energy is
localized or not, we consider  the difference of the mode densities of TP+TAP
minus  P+AP in the trivial sector\footnote{ 
In the P+AP sector there is no localized boundary energy, and the delocalized boundary energy density is proportional 
to $1/L^2$. Hence, from P+AP we get no boundary energy at all, but the reason we subtract it
from the TP+TAP is to make the result convergent. P+AP really defines the energy of the trivial vacuum.
}.  One computes for the density difference of continuum modes with  TP+TAP
versus P+AP boundary conditions\footnote{ First, we find the proper
normalization for the continuum modes. From (\ref{a38}) one finds for TP and TAP conditions in the trivial sector 
$kL=\pi n$ and then (\ref{a38}) yields $a$, and (\ref{fla1}) and (\ref{fla2}) yield $\psi_1$ and $\psi_2$
$$
\psi_1 = A \left\{ 
\left[ e^{-i \frac{\theta}{2}}+ (-1)^n  \right] e^{ikx} + \left[ e^{i
\frac{\theta}{2}}- (-1)^n
\right] e^{-ikx} 
\right\}
$$
$$
\psi_2 = A \left\{ 
\left[ e^{-i \frac{\theta}{2}}+ (-1)^n  \right]
e^{i(kx+\frac{\theta}{2})} -
\left[ e^{i
\frac{\theta}{2}}- (-1)^n  \right] e^{-i(kx+\frac{\theta}{2})} 
\right\}
$$ with $k=\frac{\pi n}{L}$ and 
$\cos \frac{\theta}{2} = - \frac{k}{\omega}$, $\sin \frac{\theta}{2} = 
\frac{m}{\omega}$. Here $A$ is a constant which we will fix for the
normalization. For the absolute values one obtains
$$ | \psi_1 |^2 =  | A|^2  \left\{   4 + 2 \cos [ 2kx - \theta ] - 2 \cos [2kx]
\right\}
$$ $$ | \psi_2 |^2 =  | A|^2  \left\{   4 + 2 \cos [ 2kx + \theta ] - 2
\cos [2kx]
\right\}
$$ and the density of the n$^{\rm th}$ mode, normalized to unity,  is 
$$ | \psi_1 |^2+| \psi_2 |^2 = \frac{1}{L} \{  1 + \frac{1}{2} \cos
[2kx] \left(
\cos \theta -1
\right)  \}
$$   Using also the expression for the zero mode  (\ref{mody1}), one
gets for the
total density in the trivial sector with TP and TAP conditions
$$ \rho_{TP}+\rho_{TAP} = 2 \times ( | \psi_{I} |^2 + |\psi_{II}|^2 )+ 2 \times
\sum_n [ | \psi_{n,1} |^2 + |\psi_{n,2}|^2 ]
$$
$$ = 2 \times  m \frac{e^{-2mx}+ e^{-2m(L-x)}}{1-e^{-2mL}}+ \frac{1}{L}
\sum_{n=1}^N \left[ 2 +
\cos(2 k x)  ( \cos  \theta
 -1) \right]
$$ where the first term stands for the two $\omega=0$ solutions.  For the P and AP conditions in the trivial sector we get
simply $\rho_P + \rho_{AP}= \frac{1}{L} \left( 1+ \sum_{n=1}^N 2 \right)$. Neglecting
$e^{-mL}$ terms and identifying $x=0$ with $x=L$ as the boundary at $x=0$ for $-L/2 \le x \le L/2$, we get for the
difference of
TP+TAP and P+AP
$$ \{ \rho_{TP} + \rho_{TAP} \} - \{ \rho_{P}+ \rho_{AP} \}= 2 m e^{-2m|x|}   -\frac{1}{L} + \frac{1}{L}
\sum_{n=1}^{N} 
\frac{-2 m^2 \cos(2 \frac{\pi n }{L} x)}{\left(\frac{\pi n }{L}
\right)^2+ m^2}
$$
$$ =2 m e^{-2m|x|}   -\frac{1}{L} + \frac{1}{2} \left[ \frac{2}{L} - \int_{-
\infty}^\infty
\frac{dk}{\pi} \frac{2 m^2 \cos(2kx)}{k^2+m^2} \right]
$$
$$ = 2 m e^{-2m|x|} - m e^{-2m|x|} = m e^{-2mx}.
$$  }
\be
\frac{1}{2} \left[ \{ \rho_{TP}(x)+\rho_{TAP}(x) \}_{cont} - \{ \rho_{P}(x)+\rho_{AP}(x) \}_{cont} \right] = - \int_{0}^{\infty}
\frac{dk}{\pi} \frac{ m^2 \cos(2kx)}{k^2+m^2}  =  - \frac{m}{2}
e^{-2m|x|} \ \ ,
\ee so that the mode density is localized around the boundary, and leads
to a net
mode number shift $\delta n =-\frac{1}{2}$.  This computation for  Majorana
fermions in the trivial sector is equivalent to the result of   Jackiw
and Rebbi
\cite{jackiw} for Dirac fermions in the presence of a kink.  Indeed, a little
thought shows that fermions in the trivial sector `feel' the twisted periodic
boundary conditions as equivalent to a kink (or antikink) of zero width.

Instead of the mode density, we may also compute the energy density.
Using 
$\int_0^\infty \frac{\cos(ak) dk}{\sqrt{k^2+m^2}}={\rm K}_0 (am)$ 
(where ${\rm
K}_0(x)$ is the modified Bessel function) to obtain for the energy density
\be
\epsilon_{TP+TAP}(x)- \epsilon_{P+AP}(x) = - \hbar\frac{m^2}{\pi} {\rm
K}_0(2m|x|) \ \ ,
\ee which is also localized around the boundary, and leads to a net boundary
energy for twisted periodic boundary conditions in the trivial sector, $\delta
M_{bound}=\hbar m/4$. This proves that there is no delocalized energy in
(\ref{po5}) and (\ref{po6}), in complete accord with the principle of cluster
decomposition. 

The average of (\ref{po1}-\ref{po4}) or (\ref{po5}-\ref{po6}) gives
\be \frac{1}{4} \left[ M^{(1)}_f(P) + M^{(1)}_f(AP) +M^{(1)}_f(TP) +
M^{(1)}_f({TAP}) \right] = M^{(1)}_f
\label{fek2}
\ee Adding this result to (\ref{bocont}), one recovers the correct
result for the
susy kink
$M^{(1)}_s$, namely (\ref{ura1}).

With these results one may address the discrepancy for $M^{(1)}$
obtained by mode
regularization between
\cite{rebhan} and the accepted value (\ref{ura1}). In \cite{rebhan} eq.
(60) the
authors computed the one-loop correction to the energy using mode regularization
for fixed periodic boundary conditions.  Thus they should have obtained
$M^{(1)}+\hbar m/4$ (where the latter term is the localized boundary
energy for
periodic conditions in the kink sector, as indicated in (46)), and they
did. 
Their calculation was a correct application of mode regularization, but
gave the
total effect of changing from trivial to kink background with fixed periodic
boundary conditions, including the boundary energy which is not part of the
localized quantum correction to the mass of the kink.  

 This brings up another question: Why do the methods of \cite{misha}  (where
$\frac{d}{dm} \sum \omega$ was first evaluated) and of
\cite{we} (energy cut-off using a smooth interpolating function)  get the
accepted answer for the mass of the kink, even though in these methods no information is used about the bottom
continuum modes, as in mode regularization?  Answer: The
boundary condition in these works do not change the density of states.  Therefore, a formula depending only on the
density of
states through the phase shift (as \cite{misha}  and 
\cite{we} do) will give an answer for the mass independent of the boundary
conditions. Because that formula agrees with the result from the
$\bar{K}K$ 
system, which has no boundary energy, it should be correct regardless of boundary
conditions.  What is lost by these methods, and this might well be
described as
an advantage, is the calculation of the total energy, including the boundary
energy. Clearly, this type of regulation gives correctly the {\bf local} energy
density associated with the boundary, if there is such, because the
local density
is insensitive to $O(1/L)$ contributions from the bottom of the
spectrum.  By
conservation of energy, therefore, what it does not necessarily give
correctly is
the delocalized energy associated with the boundary.  Put differently, in this type
of scheme
there is no reliable information about the delocalized energy, but there is
reliable information about the mass of the  kink, embodied in a single, global
integral.  By invoking the principle of cluster decomposition, with its
implication that in a correct calculation there cannot be any
delocalized energy,
one may circumvent even the one disadvantage of these schemes.  However,
as we
have seen, with mode regulation one may check the principle directly.

\subsection{The Z$_2$ gauge symmetry}

It still remains to show that a single solution with $\omega=0$ for
fermions in a
given sector with given boundary conditions
 does not correspond to any degree of freedom at all, and also to
discuss  the
effect of such a solution on the Hilbert space.   In the mode expansion
of the
Majorana field, the coefficient $c_0$ of the zero mode  is a single, idempotent
Hermitian operator.  This follows from the equal-time canonical anticommutation
relations.  The ground state may be chosen as an eigenstate\footnote{
At first sight it may seem strange to have a ground state which is half fermionic and half bosonic, but in
2 dimensions there is less distinction between fermions and bosons. 
} of
$c_0$, so
$| ground \rangle = \frac{1}{2} (1+c_0) | \Omega \rangle$. 
Consequently, all
states in the Hilbert space may be obtained by the action of local
operators on
$| ground \rangle$.  No such operator would connect
$\frac{1}{2} (1+c_0) |  \Omega \rangle$ with 
$\frac{1}{2} (1-c_0) |  \Omega \rangle$.  For $c_0$ this is true by construction,
but more complicated operators either have one factor $c_0$ or no factor $c_0$,
and in both cases one never leaves the half of the Hilbert space one is
in. The
other half of the Hilbert space is a copy of the first under the action
of the
discrete Z$_2$ symmetry which maps fermion fields $\psi$ into
$-\psi.$  The Z$_2$ symmetry is actually a discrete gauge symmetry
because it
leaves all observables ( expectation values of operators which contain
an even
number of fermion fields ) invariant, just as in quantum mechanics phase factors
of a continuous $U(1)$ symmetry multiplying state vectors are not 
observable. 

The Z$_2$ symmetry $\psi \to - \psi$ is hidden:  That is, the kink
ground state
$(1+c_0)|\Omega \rangle$ is not manifestly invariant under it ($c_0$ is mapped
into $-c_0$ under Z$_2$).  A better way of defining the ground state
might be to
say that it consists of a set with the two elements
$(1+c_0) | \Omega \rangle$ and $(1-c_0) | \Omega \rangle$. The state
$(1+c_0) |
\Omega \rangle$ is then simply a representative.  Clearly, with this definition
the ground state is Z$_2$ gauge invariant and unique. On the other hand,
it has
been observed by Ritz et al. \cite{ritz} that the ground state is not annihilated
by the (linearized) supersymmetry generator $Q_2$. Rather it is mapped into itself,
because 
$Q_2$ is proportional to $c_0$\footnote{ Using $Q = \int [ \partial_\mu \phi
\gamma^\mu \gamma^0 \psi + U \gamma^0 \psi ] dx$ one finds
$Q_2 \sim \int (\partial_x \phi_K) \psi_1 dx$. Since $\partial_x \phi_K$ is
proportional to the zero mode in $\psi_1$, see (\ref{uf10}), and
orthogonal to
the nonzero modes, one finds $Q_2 \sim c_0$. } .  Thus, half of the supersymmetry is
spontaneously broken, as has long been known, but at the same time the
unbreakable Z$_2$ gauge symmetry only is hidden, i.e., not manifest.

The necessity of averaging over more than one set of boundary conditions implies
a refinement of the assertion by Shifman et al. \cite{shifman} that boundary
conditions are unimportant if one computes the energy by calculating a regulated,
renormalized energy density and integrating this density only in the
region of
the kink.   One might expect the error in this calculation to be exponentially
small, associated with exponential localization of the boundary energy. As
discussed earlier, with the appropriate averaging over boundary
conditions   the
total delocalized  energy vanishes, which is clear because for the $\bar{K}K$
system there is no delocalized energy.
 Explicit calculation shows that the difference in the trivial sector
between the
AP and P contributions to the delocalized energy (the only kind there is
in this
case), is of order
$1/L$. Thus one can indeed forget about delocalized energy from the trivial
sector. However, the difference
$m/2$ between sums for TP and TAP conditions in the kink sector
represents a
translationally invariant contribution to the energy,  which would imply a
spurious finite shift in the energy of the  kink.  

Thus the assertion in
\cite{shifman} that the kink mass may be calculated by integrating only over,
say, the half-space surrounding the kink, and staying well away from the
boundary, indeed is correct, but with the proviso that boundary
conditions which
provide infrared regulation in this calculation must respect the Z$_2$ gauge
symmetry:  All of the boundary conditions considered above may be
visualized as
jump conditions for wavefunctions defined on a circle.  The effect of introducing
a Z$_2$ flux through that circle when a kink is present would be to interchange
TP and TAP conditions, but we have seen that there is one less mode just above
the mass threshold for TAP than for TP.  Because the coupling to this
flux is a
discrete form of a continuous gauge symmetry, a change in the number of states
would in the continous version for complex fermion fields correspond to the
abrupt disappearance of a unit of conserved charge, constituting an anomalous
violation of the gauge invariance.  To prevent such an anomaly, it is necessary
and sufficient to use a regulation which preserves the gauge invariance, namely,
describing the system as an incoherent, equal, superposition of TP and TAP
boundary conditions.  It is for this reason that in the discussion of
the energy
sums involving four different boundary conditions we have further
bundled the
sums into pairs, (TP + TAP) (essential bundling) and (P + AP) (allowed
but not
essential bundling).

There is another Z$_2$ symmetry of the action, the transformation
$\phi\to -\phi$, and simultaneously the twist $\psi\to\pm\sigma_1\psi$. This
transformation imposed as a jump condition is locally invisible, but globally
changes a system whose lowest state is the trivial vacuum to one whose lowest
state is a kink, or more precisely, a half-sphaleron:  For a circle of
circumference $2L$  with a sphaleron in metastable equilibrium, join any two
points, making a circle of circumference $L$. The matching or boundary conditions
on the half-sphaleron are precisely those of the second Z$_2$ symmetry, and
because the lowest energy configuration in this domain must be a kink
(or equally
well, an antikink), slightly squeezed because it is on a circle, it is obvious
that the field is precisely that of half the sphaleron.  The main, if
not the
only difference, from the full sphaleron is that there is no
instability, because
there is no possibility of kink-antikink annihilation in the presence of
the jump
condition.  As with the sphaleron, if $L$ is too small, then the lowest solution
is simply $\phi_{\rm classical} = 0$.  However, for larger $L$, the
half-sphaleron becomes absolutely stable.  It is, however, not a BPS solution
even at the classical level, because the BPS bound can only be saturated
on the
infinite line\footnote{ The BPS equation $\partial_x \phi + U(\phi) = 0
$ at the point where $\phi$ is maximal (and thus
$\partial_x \phi = 0$ ) requires that $U(\phi)=0$, but then the only
solution is
$\phi=m/\sqrt{2
\lambda}$. } .  

This discussion complements a recent analysis by Binosi et al.
\cite{binosi}  of solitons with winding number
$\pm 1$ in an $N=2$ supersymmetric theory with a potential depending on
a complex
$\phi$ and periodic in $\Re \phi$. It is clear that there is quantum tunneling
between soliton and antisoliton in both cases.  However, their $N=2$
soliton is
quantum unstable but classically saturates the BPS bound, whereas our
kink is
quantum stable but already violates the BPS bound at the classical
level.  A
tentative conclusion from these two examples is that tunneling between soliton
and antisoliton is likely to be the most generic feature for such systems.

We now are in a position to address an issue glossed over above:  How general
is the statement that the
methods of
\cite{rebhan} and \cite{misha} are independent of boundary conditions?
One conspicuous case for which these methods do not
work, as mentioned already in the introduction, is that of `supersymmetric'
boundary conditions
$\phi(0)=\psi_1(0)=\phi(L)=\psi_1(L)=0$.  There are two reasons for this failure.
First, the mentioned Dirac spinor conditions are equivalent to energy-dependent
boundary conditions in a Schr\"{o}dinger equation, and hence do change the
density of states.  Secondly, this choice of boundary conditions obviously
violates the second Z$_2$ symmetry, under which $\psi_1(0)=\psi_1(L)=0$ in
the kink sector $\to
\psi_1(0)=\psi_2(L)=0$ in the trivial sector.  To average over a set of fixed
boundary conditions while respecting the  Z$_2$ symmetry, one must have both
these choices in both sectors.  Indeed, doing so will remove the logarithmic
divergence mentioned earlier, and also will reproduce the change in
density of
states between trivial and kink sectors due to the kink alone. (The details for these boundary conditions
will be worked out explicitly elsewhere \cite{future}).
Not surprisingly, the logarithmically divergent energy associated with violating
the second Z$_2$ symmetry corresponds to a delocalized energy density, which
gives another reason why such a choice of boundary condition should not be
allowed.  That this energy is delocalized is evident from the fact that the
change in density of states comes from the energy dependence of the boundary
conditions, which simply spaces continuum levels differently `by fiat',
so that
there is no position where the associated energy shift should be localized.

Should both these discrete symmetries be
considered gauge symmetries?  As we have seen already, the second Z$_2$ is
similar to the first in that each one when applied to a jump condition
has global
implications.  The first introduces a half-quantum of flux through the
circle on
which the fields are defined, while the second converts the sector with kink
number 0 mod 2 into the sector with kink number 1 mod 2.  Nevertheless, while
globally significant, both these jump conditions may be applied at any
point, and
there will be no local consequences at that point. However, there are two
important reasons not to describe the second Z$_2$ as also a gauge symmetry.
1)  If one goes from real fields to complex fields, so that  Z$_2$
becomes U(1),
then the first corresponds to standard gauge coupling, but the second would
correspond to axial gauge coupling.  It is well known that this U(1) $\times$
U(1) is an anomalous theory.  2)  In the sector with the second  Z$_2$ jump
condition, as mentioned earlier it is possible to describe paths in field
configuration, with action proportional to the circumference of the circle,
connecting kink and antikink.  This would make no sense if kink and antikink
were identical, as would be implied by treating the second  Z$_2$ as a gauge
symmetry.  One may see this also by considering the sphaleron
configuration, in
which there may be a convention used to define the kink, but the distinction
between kink and antikink is clear from the fact that they can
annihilate each
other.  Indeed, the same point is manifest already for the vacuum configuration
$\eta = \pm{\rm constant}$.  For a finite-circumference ring, there will be
tunneling between the positive and negative values, leading to two nearly
degenerate ground states which are equal superpositions of the two
values. 
Only for infinite circumference do we have the thermodynamic limit in which
spontaneous symmetry breaking occurs, and the two values are completely
independent.

\section{Conclusions }

Mode regularization, i.e.,  the simple prescription that one subtracts vacuum
energies for the same number of modes with and without some background, differs
from many other regularization schemes in that the cutoff parameter need
not be
averaged over some continuous weight function  which goes from unity at low
energies to zero at high  energies, as is necessary in particular for
energy or
momentum cutoff \cite{we}.  This attractive feature gives a strong
incentive to
investigate whether the scheme is universally applicable. Here we have  studied
mode regularization for the case of the kink in (1+1) dimensions, including
Majorana fermions. There is a well-known subtlety in counting boson zero
 modes, that zero-frequency modes must be expressed as collective
coordinates, so
that for each coordinate there is a conjugate momentum, giving rise to
raising and
lowering operators just like those for nonzero frequencies.   Thus the two
bosonic zero modes of a widely separated kink-antikink system become two
collective coordinates, and they correspond to two pairs of $(P,X)$ variables,
counting as two degrees of freedom. This makes sense because it keeps
the total
number of modes constant as the corresponding squared frequency
$\omega^2$ goes to or through zero.

For fermions the situation is less familiar.  By studying the problem in three
closely related systems, kink-antikink, sphaleron (i.e., kink and antikink
symmetrically placed on a circle), and isolated kink, we find a very different
behavior from that for bosons, namely, the number  of $\omega \sim 0$
degrees of
freedom (one) in the
$\bar{K}K$ system is half the number (two) of fermionic $\omega \sim 0 $
solutions:  The two zero modes in the Dirac equation for the fermions together
give one annihilation and one creation  operator in the expansion of the fermion
field, hence one degree of freedom.  

This suggests that for an isolated kink, there is one boson zero mode
degree of
freedom,  but only half a fermion zero mode degree of freedom.  This
half is
interpreted as being due to different boundary conditions for which the energies
must be averaged to give the correct mass shift.  Two of the boundary conditions
(periodic twisted and antiperiodic twisted) give a single zero mode in
the Dirac
equation, and the hermitian coefficient $c_0$ of this mode function in
the fermion
field leaves the ground state invariant.   There is no doubling of the Hilbert
space due to this $c_0$ because the states
$(1+c_0)|\Omega \rangle $ and $(1-c_0)|\Omega \rangle $  are equivalent
under the
Z$_2$ symmetry $\psi \to -\psi$. 

The conclusion that the ground state is an eigenstate of a fermionic operator 
is at first thought puzzling.  It not only violates intuition based on
widespread experience, but also appears to contradict the well-known superselection rule forbidding
coherent superposition of states with even and odd fermion number. 
Although Majorana fermions do not carry an additive, conserved fermion
number $F$, still the fermion field anticommutes with the ${ \rm Z_2}$
factor $(-1)^F$.  However, in one space dimension the distinction
between bosons and fermions is not pronounced, because there is no spin
connected to the statistics; for example, one can
bosonise fermions in string theory. The susy multiplet also is unusual:
it contains 2 states for the non-BPS case \cite{we}, but only one state for the BPS 
case, as discussed in \cite{shifman} and this article.  Thus, the state $(1+c_0)|\Omega \rangle $ is not an
exception to the superselection rule, but rather a unique and unexpected illustration of that rule.

The Z$_2$ symmetry $\psi \to -\psi$  is actually a gauge 
symmetry because it leaves all possible observables invariant. That is,
there is
no conceivable field which could be added to the Lagrangian as a
perturbation and
would give a coupling not invariant under $\psi\to -\psi$.  Thus there
is no
zero-frequency fermionic degree of freedom for these boundary
conditions, and
consequently one must include one more fermion than boson continuum mode
in the
mode regularization if one wishes to consider only TP and TAP boundary conditions.
The two other boundary conditions (periodic and antiperiodic) each give
in the
kink sector two fermionic solutions with
$\omega=0$ (P) or $\omega\sim 0$ (AP).  Their coefficients yield one annihilation
and one creation operator in the Dirac field expansion, and hence one
corresponding fermionic degree of freedom, half localized at the kink
and half
localized at the boundary.

Evidently, the average of the sums with different boundary conditions is 
equivalent to the loss of half a fermionic degree of freedom.  This half clearly
is related to the half fermion charge found by Jackiw and Rebbi \cite{jackiw}
when they considered Dirac fermions in the presence of a kink. For a Majorana
fermion the half degree of freedom of the kink system was initially interpreted
as one degree of freedom in the $\bar{K} K$ system which was shared half
by the
kink and half by the antikink.
 The nonlocal character of this fermionic degree of freedom for the
${\rm K
\bar{K}}$ system is still present for plain periodic boundary conditions
on the
kink alone:  The degree of freedom is localized half at the kink (by $\psi_1$)
and half at the boundary (by $\psi_2$).  For our Majorana fermion we can
interpret this fractionalization of degree of freedom as follows:
Suppose one
starts in the trivial vacuum with periodic boundary conditions, and one
starts to
rotate the right-hand half of the  constant background field $\phi$ by a
Goldstone-Wilczek chiral rotation \cite{gw}. This rotation produces a
current and
changes the trivial vacuum to the kink vacuum. The periodic boundary
condition in
the kink sector stops the current, and half a degree of freedom is  accumulated
at the boundary. In the analysis of Jackiw and Rebbi \cite{jackiw} the boundary
was moved to infinity and thus they only found half a charge around the
kink. 

The results summarized here lead to a well-defined procedure for
applying mode
regularization to a system in which the boundary conditions naturally change
between one sector and another:  In the difference of sums, require that the
terms of highest energy are matched in such a way that there is no contribution
to the quantum energy shift linearly divergent with the maximum energy $\Lambda$.
Having thus matched the sums `at the top' one may count down to the
bottom, and
compare the number of modes in each sum.  For fixed boundary conditions, this
procedure is equivalent to the usual mode-counting prescription. 
However, when
the boundary conditions change, as from P+AP to TP+TAP, then the number
of modes
can change.   
(For the P, AP cases in the trivial sector and the TP case in the kink sector there are equal numbers of nonzero modes, 
but for the TAP case in the kink sector there is one less zero mode.)
Because the number goes down by 1 for TAP but zero for TP, the
average loss is 1/2.  Thus, one important conclusion from our work is that
there is a natural generalization of mode regularization to the case when
boundary conditions are not fixed.  This may be useful in other contexts.

The nonlocality of one fermionic degree of freedom clearly must be an essential
feature of a theory where there is only one unpaired fermion state
localized at a
soliton.  For the case of locally invisible boundary conditions both in the
trivial sector (P + AP) and in the kink sector (TP + TAP), the half
charge is
even more ethereal.  It simply evaporates under the change in boundary
conditions, as a consequence of the chiral anomaly. 

The nonlocality is surprising because it appears to violate the
principle of
cluster decomposition.  However, Majorana fermion charge is not an observable,
and all vacuum expectation values for observable fields still obey the
principle.   Of course, if the fermion field carried an observable
charge, such
as fermion charge for a Dirac fermion, then the half would become a localized
eigenvalue, as in the case analyzed by Jackiw and Rebbi.  Thus, the unadorned
Majorana fermion interacting with the kink is a kind of ``square root''
of the
Dirac fermion, still manifesting the Jackiw-Rebbi half charge, but in a
delocalized form,  whose precise specification depends on what might otherwise
have seemed an arbitrary choice of boundary conditions.  

The combined thrust of all the recent works is to show that new aspects
continue to
appear in understanding a prototype soliton, the kink in 1+1 dimensions,
with a strong hint that the methods sharpened in this theoretical laboratory
are likely to have broader application.   

{\bf Acknowledgements:} We thank N. Graham, N. Manton, A. Rebhan, M. Ro\v cek,
J. Schonfeld, A. Sen, M. Shifman, W. Siegel, M. Stephanov and  A. Vainshtein  for
discussions. These results were presented at a meeting celebrating 30
years  of
supersymmetry  in Minnesota. A new paper by Losev, Shifman and Vainshtein,
\cite{losev},  has discussed the multiplet structure in $N=1$ supersymmetric
models. Their result seem to be complementary and compatible with ours.


\begin{thebibliography}{99}
\bibitem{raj} R.Rajaraman, {\it Solitons and instantons}, Elsevier, (1996).
\bibitem{rebhan} A.Rebhan and P. van Nieuwenhuizen, {\it No saturation
of the
quantum Bogomolnyi bound by two-dimensional N=1 supersymmetric solitons},
Nucl. Phys. {\bf B508}, 449 (1997).
\bibitem{misha} H. Nastase, M. Stephanov, P. van Nieuwenhuizen and A. Rebhan,
 {\it Topological boundary conditions, the BPS bound, and elimination of
ambiguities in the quantum mass of solitons}, Nucl.Phys. {\bf B542}, 471 (1999)
, hep-th/9802074 
\bibitem{we} A. Litvintsev and P. van Nieuwenhuizen, {\it Once more on
the BPS
bound for the susy kink}, hep-th/0010051
\bibitem{graham} N.Graham and R.L.Jaffe, {\it Energy, Central Charge,
and the BPS
Bound for 1+1 Dimensional Supersymmetric Solitons}, Nucl. Phys. {\bf B544},
432 (1999), hep-th/9808140v3;  N. Graham and R.L. Jaffe, {\it Fermionic one
loop corrections to soliton energies in (1+1)-dimensions}, Nucl. Phys.{\bf
B549}, 516 (1999), hep-th/9901023;  N. Graham and R.L. Jaffe, {\it Unambiguous
One-Loop Quantum Energies of  1+1 Dimensional Bosonic Field
Configurations}, Phys. Lett. {\bf B435}, 145 (1998), hep-th/9805150;   E. Farhi,
N. Graham, R.L. Jaffe, and H. Weigel {\it Heavy Fermion Stabilization of Solitons
in 1+1 Dimensions}, hep-th/0003144
\bibitem{shifman} M. Shifman, A. Vainshtein and M. Voloshin, {\it
Anomaly and
Quantum Corrections to Solitons in Two-Dimensional Theories with Minimal
Supersymmetry}, Phys. Rev. D {\bf59}, 45016 (1999), hep-th/9810068v2
\bibitem{dunne} G.V. Dunne, {\it Derivative Expansion and Soliton Masses},
hep-th/9907208 
\bibitem{schonfeld} J.F. Schonfeld, {\it Soliton Masses in Supersymmetric
    Theories}, Nucl. Phys. {\bf B161}, 125 (1979).
\bibitem{WO}  E. Witten and D. Olive, {\it Supersymmetry Algebras That Include
    Topological Charges}
    Phys. Lett. {\bf B78}, 97 (1978).
\bibitem{manton} N.S. Manton and T.M. Samols, {\it Sphalerons on a circle},
Phys. Lett. {\bf B207},  179 (1988).
\bibitem{korn} G.A. Korn and T.M. Korn, {\it Mathematical handbook}, McGraw-Hill,
(1968); E.T. Whittaker and G.N. Watson, {\it A course of modern
analysis},  Camb.
Univ. Press, chapter XXII (1962).
\bibitem{gw} J. Goldstone and F. Wilczek, {\it Fractional Quantum
Numbers on
Solitons}, Phys. Rev. Lett. {\bf 47},  986 (1981).
\bibitem{jackiw} R.Jackiw and C.Rebbi, {\it Solitons with Fermion Number 1/2},
Phys. Rev.  D {\bf 13},  3398 (1976).
\bibitem{ritz} A. Ritz, M. Shifman, A. Vainshtein and M. Voloshin, {\it Marginal
Stability and the Metamorphosis of BPS States}, hep-th/0006028
\bibitem{binosi} D. Binosi, M. Shifman and T. ter Veldhuis, {\it Leaving
the BPS
Bound: Tunneling of Classically Saturated Solitons}, hep-th/0006026
\bibitem{future} A. S. Goldhaber, A. Litvintsev and P. van Nieuwenhuizen, to be publish
ed
\bibitem{losev}  A. Losev, M. Shifman and  A. Vainshtein, {\it Single State
Supermultiplet in 1+1 Dimensions}, hep-th/0011027

\end{thebibliography}
\end{document}